\documentclass[%
 aip,
 jmp,%
 amsmath,amssymb,
reprint,%
]{revtex4-1}

\usepackage[colorlinks,linkcolor=blue,citecolor=blue,urlcolor=black,bookmarks=false,hypertexnames=true]{hyperref} 

\usepackage{amsmath}
\usepackage{xspace}
\usepackage{multirow}
\usepackage{latexsym}
\usepackage{braket}
\usepackage{cleveref}
\usepackage{enumerate}
\usepackage{mathrsfs}
\usepackage{enumitem}
\usepackage{color}
\usepackage[version=3]{mhchem}
\usepackage{caption,setspace}
\usepackage{graphicx}
\usepackage{subfigure}
\usepackage{threeparttable}
\usepackage{url}

\allowdisplaybreaks
\usepackage{array}
\newcolumntype{L}[1]{>{\raggedright\let\newline\\\arraybackslash\hspace{0pt}}m{#1}}
\newcolumntype{C}[1]{>{\centering\let\newline\\\arraybackslash\hspace{0pt}}m{#1}}
\newcolumntype{R}[1]{>{\raggedleft\let\newline\\\arraybackslash\hspace{0pt}}m{#1}}

\newcommand{\h}[2]{h_{{#2}}^{{#1}}}
\newcommand{\f}[2]{f_{{#2}}^{{#1}}}

\renewcommand{\v}[2]{{v}_{{#2}}^{{#1}}}
\renewcommand{\t}[2]{{t}_{{#2}}^{{#1}}}
\renewcommand{\c}[1]{a^\dagger_{#1}}
\renewcommand{\a}[1]{a_{#1}}

\newcommand{\e}[1]{\ensuremath{\varepsilon_{#1}}}

\newcommand{\SC}{\ensuremath{\Delta S^{2}}\xspace}

\crefname{figure}{Figure}{Figures}
\crefname{table}{Table}{Tables}
\crefname{equation}{Eq.}{Eqs.}
\crefname{section}{Section}{Sections}

\renewcommand{\t}[2]{{t}_{{#1}}^{{#2}}}
\renewcommand{\a}[1]{a^{\ }_{#1}}

\newcommand*{\mae}{$\mathrm{MAE}$\xspace}
\newcommand*{\std}{$\mathrm{STDV}$\xspace}

\begin{document}

\author{Terrence~L.~Stahl}
\affiliation{Department of Chemistry and Biochemistry, The Ohio State University, Columbus, Ohio 43210, USA}

\author{Samragni~Banerjee}
\affiliation{Department of Chemistry and Biochemistry, The Ohio State University, Columbus, Ohio 43210, USA}

\author{Alexander~Yu.~Sokolov}
\email{sokolov.8@osu.edu}
\affiliation{Department of Chemistry and Biochemistry, The Ohio State University, Columbus, Ohio 43210, USA}

\title{
\color{blue}
    Quantifying and reducing spin contamination in algebraic diagrammatic construction theory of charged excitations
\vspace{0.25cm}
}

\begin{abstract}
	Algebraic diagrammatic construction (ADC) theory is a computationally efficient and accurate approach for simulating electronic excitations in chemical systems.
	However, for the simulations of excited states in molecules with unpaired electrons the performance of ADC methods can be affected by the spin contamination in unrestricted Hartree--Fock (UHF) reference wavefunctions. 
	In this work, we benchmark the accuracy of ADC methods for electron attachment and ionization of open-shell molecules with the UHF reference orbitals (EA/IP-ADC/UHF) and develop an approach to quantify the spin contamination in the charged excited states.
	Following this assessment, we demonstrate that the spin contamination can be reduced by combining EA/IP-ADC with the reference orbitals from restricted open-shell Hartree--Fock (ROHF) or orbital-optimized M$\o$ller--Plesset perturbation (OMP) theories.
	Our numerical results demonstrate that for open-shell systems with strong spin contamination in the UHF reference the third-order EA/IP-ADC methods with the ROHF or OMP reference orbitals are similar in accuracy to equation-of-motion coupled cluster theory with single and double excitations.	
\end{abstract}

\titlepage

\maketitle

\section{Introduction}
\label{sec:intro}

Charged excitations are an important class of light-matter interactions that result in a generation of free charge carriers (electrons or holes) in a chemical system. 
Theoretical simulations of charged excitations find applications in determining the redox properties of molecules, understanding the electronic structure of materials, interpreting the photoelectron spectra, and elucidating the mechanisms of photoredox catalytic reactions and genetic damage of biomolecules.\cite{seki:1989p255,li:2018p35108,d:2005p11,perdew:2017p2801,dittmer:2019p9303,tripathi:2019p10131,schultz:2014p343}
However, accurate modeling of charged excitations is very challenging as it requires an accurate description of electron correlation, orbital relaxation, and charge localization effects. 

Of particular interest are charged excitations of open-shell molecules with unpaired electrons in the ground electronic state.
These excitations are common in atmospheric and combustion chemistry,\cite{wallington:1996p18116,aloisio:2000p825,tyndall:2001p12157,glowacki:2010p3836,narendrapurapu:2011p14209,sun:2012p8148} but are also key to many important reactions in organic synthesis.\cite{ramaiah:1987p3541,bertrand:1994p257,glowacki:2010p3836,denes:2014p2587,khamrai:2021p1,lee:2021p14081}
Simulating charged excitations in open-shell molecules presents new challenges, such as an accurate description of electronic spin states and an increased importance of electron correlation effects.
In particular, inadequate description of electronic spin in the ground or charged excited states can lead to spin contamination that can significantly affect the performance of approximate electronic structure theories. 
Spin contamination can be mitigated using a variety of theoretical approaches including coupled cluster theory (in its state-specific\cite{Crawford:2000p33,Shavitt:2009} or equation-of-motion\cite{Nooijen:1992p55,stanton:1993p7029,Nooijen:1993p15,Nooijen:1995p1681,Crawford:2000p33,Krylov:2008p433,Shavitt:2009} formulation), orbital-optimized methods,\cite{scuseria:1987p354,krylov:1998p10669,Sherrill:1998p4171,gwaltney:2000p3548,lochan:2007p164101,bozkaya:2011p224103,bozkaya:2011p104103,bozkaya:2013p184103,bozkaya:2013p104116,bozkaya:2012p204114,Bozkaya:2013p54104,Sokolov:2013p024107} and multireference theories.\cite{Buenker:1974p33,Siegbahn:1980p1647,Werner:1988p5803,Mukherjee:1977p955,Evangelista:2007p024102,Datta:2011p214116,Kohn:2012p176,Datta:2012p204107,Banerjee:1978p389,Nichols:1998p293,Khrustov:2002p507,chatterjee:2019p5908,chatterjee:2020p6343,de:2022p4769}
A common problem of these approaches is a high computational cost that limits their applicability to small open-shell molecules.

An attractive alternative to conventional electronic structure methods for simulating excited states is algebraic diagrammatic construction theory (ADC).\cite{Schirmer:1982p2395,Schirmer:1991p4647,Mertins:1996p2140,Schirmer:2004p11449,Dreuw:2014p82}
ADC offers a framework of computationally efficient and accurate approximations that can compute many excited states incorporating the description of electron correlation effects in a single calculation.
Although the ADC methods for charged excited states have been formulated several decades ago,\cite{Schirmer:1983p1237,Schirmer:1998p4734,Trofimov:2005p144115} their development and efficient computer implementation have received a renewed interest recently.\cite{Trofimov:2011p77,Schneider:2015p144103,Dempwolff:2019p064108,banerjee:2019p224112,dempwolff:2020p024113,dempwolff:2020p024125,Liu:2020p174109,banerjee:2021p74105,Dempwolff:2021p104117}
In particular, the non-Dyson ADC methods\cite{Schirmer:1998p4734,Trofimov:2005p144115,Dempwolff:2019p064108,banerjee:2019p224112} allow for independent calculations of electron-attached and ionized excited states with accuracy similar to that of coupled cluster theory with single and double excitations (CCSD),\cite{Crawford:2000p33,Shavitt:2009} but at a fraction of its computational cost. 
However, the applications of ADC methods for charged excitations have been primarily limited to closed-shell molecules, with questions remaining about their accuracy for systems with unpaired electrons in the ground electronic state.\cite{Dempwolff:2019p064108,banerjee:2019p224112,Dempwolff:2021p104117}
As a finite-order perturbation theory, ADC is expected to be more sensitive to spin contamination than coupled cluster theory and thus may be less accurate when the errors in electronic spin become significant.
However, a thorough study of spin contamination in the ADC calculations of charged excitations has not been reported.

In this work, we develop an approach to quantify the spin contamination in ADC calculations of charged excited states of open-shell molecules and investigate how the errors in describing the electronic spin affect the performance of ADC methods. 
We further demonstrate that the errors in charged excitation energies and spin contamination can be reduced by combining the ADC methods with the reference orbitals from restricted open-shell Hartree--Fock (ROHF)\cite{roothaan:1960p179} or $n$th-order orbital-optimized M$\o$ller--Plesset perturbation (OMP($n$))\cite{bozkaya:2013p184103,bozkaya:2013p104116,bozkaya:2011p224103,bozkaya:2011p104103,Lee:2019p244106,bertels:2019p4170,Neese:2009p3060,lochan:2007p164101,kurlancheek:2009p1223} theories.
The resulting theoretical approaches are shown to have similar accuracy to equation-of-motion CCSD for open-shell molecules with strong spin contamination in the unrestricted Hartree--Fock reference wavefunction.

\section{Theory}
\label{sec:theory}

\subsection{Algebraic diagrammatic construction theory}
\label{sec:theory:Propagator_ADC_Theory}
The central mathematical object in algebraic diagrammatic construction theory of charged excitations is the one-particle Green's function (1-GF, also known as the electron propagator) that contains information about electron affinities (EA) and ionization energies (IP) of a many-electron system.\cite{Fetter2003,Dickhoff2008,danovich:2011p377} 
In its spectral (Lehmann) representation, 1-GF is defined as
\begin{align}
\label{eqn:spectral}
\mathit{G}_{pq}(\omega) &=\sum_{n}\frac{\braket{\Psi_{0}^{N} |a_{p}| \Psi_{n}^{N+1}}\braket{\Psi_{n}^{N+1}|a_{q}^{\dagger}| \Psi_{0}^{N}}}{\omega-E_{n}^{N+1}+E_{0}^{N}} \notag \\
&+ \sum_{n}\frac{\braket{\Psi_{0}^{N} |a_{q}^{\dagger}| \Psi_{n}^{N-1}}\braket{ \Psi_{n}^{N-1}|a_{p}| \Psi_{0}^{N}}}{\omega+E_{n}^{N-1}-E_{0}^{N}} \notag \\
&= \mathit{G}_{+pq}(\omega) + \mathit{G}_{-pq}(\omega) \ ,
\end{align}
where $\mathit{p,q}$ index molecular orbitals of the system, $\omega$ is a complex frequency ($\omega \equiv \omega^{'} +\mathit{i}\eta$), $\mathit{G}_{+pq}(\omega)$ and $\mathit{G}_{-pq}(\omega)$ are the forward (EA) and backward (IP) components of the propagator, respectively.
In \cref{eqn:spectral}, $\ket{ \Psi_{0}^{N}}$ is the $N$-electron ground-state wavefunction with energy $E_{0}^{N}$, while $\ket{\Psi_{n}^{N+1}}$ and $\ket{\Psi_{n}^{N-1}}$ are the $(N+1)$- and $(N-1)$-electron excited states with energies $E_{n}^{N+1}$ and $E_{n}^{N-1}$.
In \cref{eqn:spectral}, the numerators containing the expectation values of creation ($a_{p}^{\dag}$) and destruction ($a_{p}$) operators (so-called residues) describe the probability of EA and IP transitions with energies $\omega_{+n} = E_{n}^{N+1} - E_{0}^{N}$ and $\omega_{-n} =E_{0}^{N} -  E_{n}^{N-1}$, respectively.

 \cref{eqn:spectral} can be rewritten in a matrix form for each component of 1-GF individually:
\begin{align}
\label{eqn:exactpropagator}
\bold{G}_{\pm}(\omega) = \tilde{\bold{X}}_{\pm}(\omega\mathbf{1}-\tilde{\bold{\Omega}}_{\pm})^{-1} \tilde{\bold{X}}_{\pm}^{\dagger} \ ,
\end{align}
where the diagonal matrix $\tilde{\bold{\Omega}}_{\pm}$ contains the transition energies $\tilde{\Omega}_{\pm nm}=\omega_{\pm n}\delta_{nm}$ and $\tilde{\bold{X}}_{\pm}$ is the matrix of spectroscopic (or transition) amplitudes with elements $\tilde{X}_{+pn}=\braket{\Psi_{0}^{N} |a_{p}|\Psi_{n}^{N+1}}$ or $\tilde{X}_{-qn}=\braket{\Psi_{0}^{N} |a_{q}^{\dagger}|\Psi_{n}^{N-1}}$. 
\cref{eqn:exactpropagator} provides a prescription for calculating 1-GF, but is rarely used in practice since the exact (i.e., full configuration interaction) eigenstates $\ket{\Psi_{n}^{N\pm1}}$ and excitation energies $\omega_{\pm n}$ are prohibitively expensive to compute for realistic systems and basis sets. 

Algebraic diagrammatic construction theory (ADC)\cite{Schirmer:1982p2395,Schirmer:1991p4647,Mertins:1996p2140,Schirmer:2004p11449,Dreuw:2014p82} provides a computationally efficient approach to circumvent this problem by approximating the forward and backward components of the propagator independently of each other using perturbation theory (so-called non-Dyson approach).\cite{Schirmer:1998p4734,Trofimov:2005p144115,Dempwolff:2019p064108,banerjee:2019p224112}
ADC starts by reformulating \cref{eqn:exactpropagator} into its non-diagonal form
\begin{align}
\label{eqn:non_diag}
\bold{G}_{\pm}(\omega) = \bold{T}_{\pm} ( \bold{\omega}\mathbf{1}-\bold{M}_{\pm}  )^{-1}  \bold{T}_{\pm}^{\dagger} \ ,
\end{align}
where $\bold{M}_{\pm}$ is the non-diagonal effective Hamiltonian matrix and $\bold{T}_{\pm}$ is the matrix of effective transition moments. 
The $\bold{M}_{\pm}$ and $\bold{T}_{\pm}$ matrices are expanded in a perturbative series, which is truncated at order $n$,
\begin{align}
\label{eqn:M_pt}
\bold{M}_{\pm}  &\approx \bold{M}^{(0)}_{\pm} + \bold{M}^{(1)}_{\pm} + \ldots + \bold{M}^{(n)}_{\pm} \ ,  \\
\label{eqn:T_pt}
\bold{T}_{\pm}  &\approx \bold{T}^{(0)}_{\pm} + \bold{T}^{(1)}_{\pm} + \ldots + \bold{T}^{(n)}_{\pm} \ , 
\end{align}
defining the ADC($n$) approximation for the propagator.\cite{Schirmer:1998p4734}
Diagonalizing the $\bold{M}_{\pm}$ matrix
\begin{equation}
\label{eqn:eigenvalue}
\bold{M}_{\pm}\bold{Y}_{\pm}=\bold{Y}_{\pm}\bold{\Omega}_{\pm}
\end{equation}
yields the approximate EA's or IP's, as well as the eigenvectors $\bold{Y}_{\pm}$ that can be used to compute the approximate spectroscopic amplitudes $\bold{X}_{\pm}=\bold{T}_{\pm}\bold{Y}_{\pm}$.

\subsection{ADC equations from effective Liouvillian theory}
\label{sec:theory:adc_equations}
Working equations for the matrix elements of $\bold{M}_{\pm}$ and $\bold{T}_{\pm}$ can be derived from the algebraic analysis of 1-GF,\cite{Schirmer:1982p2395} using the intermediate state representation approach,\cite{Schirmer:1991p4647,Schirmer:1998p4734,Trofimov:1999p9982,Dempwolff:2019p064108} or the formalism of effective Liouvillian theory.\cite{Prasad:1985p1287,Mukherjee:1989p257,Sokolov:2018p204113,banerjee:2019p224112,banerjee:2021p74105} 
Here, we review the equations of single-reference ADC theory derived using the effective Liouvillian approach where the ground-state wavefunction $\ket{\Psi_{0}^{N}}$ is assumed to be well-approximated by a reference Slater determinant $\ket{\Phi}$. 
Generalization of the effective Liouvillian and ADC theories to multiconfigurational reference states has been described elsewhere.\cite{Sokolov:2018p204113,chatterjee:2019p5908,chatterjee:2020p6343,de:2022p4769,mazin:2021p6152}

\begin{figure*}[t!]
	\centering
	\captionsetup{justification=raggedright,singlelinecheck=false,font=footnotesize}
	\includegraphics[scale=0.40,trim=0.0cm 0.0cm 0.0cm 0.0cm,clip]{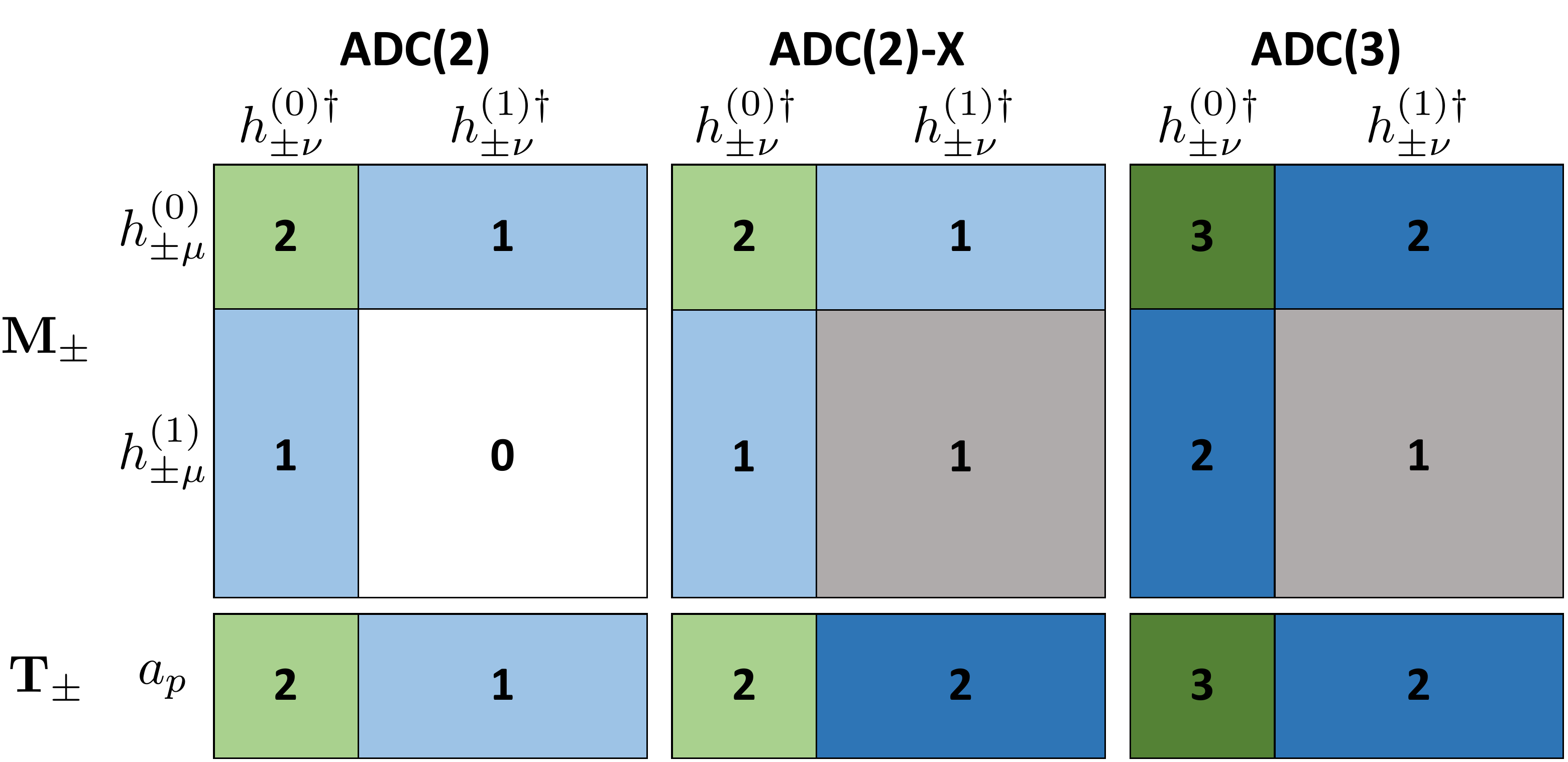}
	\caption{Perturbative structure of the $\bold{M}_{\pm}$ and $\bold{T}_{\pm}$ matrices in ADC for charged excitations.
		Numbers indicate the perturbation order to which the effective Hamiltonian $\tilde{H}$ and observable $\tilde{a}_{p}$ operators are expanded for a particular block of $\bold{M}_{\pm}$ and $\bold{T}_{\pm}$, respectively. }
	\label{fig:ADC_matrices}
\end{figure*}

In the single-reference ADC theory, the perturbative expansion in \cref{eqn:M_pt,eqn:T_pt} is generated by separating the Hamiltonian $H$ into a zeroth-order contribution
\begin{align}
	\label{eqn:h_zero}
	H^{(0)} = E_0 + \sum_{p} \e{p} \{ \c{p}\a{p} \}
\end{align}
and a perturbation
\begin{align}
	\label{eqn:perturbation}
	V = \frac{1}{4} \sum_{pqrs} \v{pq}{rs} \{ \c{p}\c{q}\a{s}\a{r} \} \ ,
\end{align}
where $E_0 = \braket{\Phi|H|\Phi}$ is the Hartree--Fock energy, $\e{p}$ are the eigenvalues of the canonical Fock matrix (so-called orbital energies)
\begin{align}
	\label{eqn:fock_matrix}
	f_p^q = \h{p}{q} + \sum_{i}^{occ} \v{pi}{qi} \ , 
\end{align}
$\h{p}{q} = \braket{p|h|q}$ and $\v{pq}{rs} = \braket{pq||rs}$ are the one-electron and antisymmetrized two-electron integrals. 
Notation $\{\ldots\}$ indicates that the creation and annihilation operators are normal-ordered with respect to the reference determinant $\ket{\Phi}$. 
Indices $p,q,r,s$ run over all spin-orbitals in a finite one-electron basis set, while $i,j,k,l$ and $a,b,c,d$ index the occupied and virtual orbitals, respectively.
We note that the ADC zeroth-order Hamiltonian is equivalent to the one used in single-reference M\o ller--Plesset perturbation theory.\cite{Moller:1934kp618}

Starting with the zeroth-order Hamiltonian in \cref{eqn:h_zero} and using the effective Liouvillian approach for deriving the ADC equations, we arrive at the expressions for the $n$th-order contributions to the $\bold{M}_{\pm}$ and $\bold{T}_{\pm}$matrix elements:
\begin{align}
	\label{eqn:M_plus}
	M_{+\mu\nu}^{(n)} &= \sum_{klm}^{k+l+m=n} \braket{\Phi|[h_{+\mu}^{(k)},[\tilde{H}^{(l)},h_{+\nu}^{(m)\dagger}]]_{+}|\Phi} \ , \\
	\label{eqn:M_minus}
	M_{-\mu\nu}^{(n)} &= \sum_{klm}^{k+l+m=n} \braket{\Phi|[h_{-\mu}^{(k)\dagger},[\tilde{H}^{(l)},h_{-\nu}^{(m)}]]_{+}|\Phi} \ , \\
	\label{eqn:T_plus}
	T_{+p\nu}^{(n)} &= \sum_{kl}^{k+l=n} \braket{\Phi|[\tilde{a}_{p}^{(k)},h_{+\nu}^{(l)\dagger}]_{+}|\Phi} \ , \\
	\label{eqn:T_minus}
	T_{-p\nu}^{(n)} &= \sum_{kl}^{k+l=n} \braket{\Phi|[\tilde{a}_{p}^{(k)},h_{-\nu}^{(l)}]_{+}|\Phi} \ .
\end{align}

Here, square brackets denote commutators ($[A,B] = AB - BA$) or anticommuators ($[A,B]_{+} = AB + BA$).
The excitation operators $ h_{\pm\mu}^{(m)\dag}$ are used to form a set of basis states $\ket{\Psi^{(m)}_{\pm\mu}} = h_{\pm\mu}^{(m)\dagger}\ket{\Phi}$ to represent the eigenstates of ($N \pm 1$)-electron system. 
For the low-order ADC approximations (up to ADC(3)) only the zeroth- ($ h_{+\mu}^{(0)\dag} = \c{a}$ and $ h_{-\mu}^{(0)\dag} = \a{i}$) and first-order ($ h_{+\mu}^{(1)\dag} = \c{b}\c{a}\a{i}$ and $ h_{-\mu}^{(1)\dag} = \c{a}\a{j}\a{i}$) excitation operators appear in the ADC equations. 
The  $\tilde{H}^{(l)}$ and $\tilde{a}_{p}^{(l)}$ operators are  $l$th-order contributions to the effective Hamiltonian $\tilde{{H}}=e^{-(T-T^{\dagger})}He^{(T-T^{\dagger})}$ and effective observable $\tilde{a}_{p}=e^{-(T-T^{\dagger})}a_{p}e^{(T-T^{\dagger})}$ operators, respectively.

For each block of $\bold{M}_{\pm}$ and $\bold{T}_{\pm}$  defined by a pair of excitation operators $ h_{\pm\mu}^{(m)\dag}$, the effective operators $\tilde{H}$ and $\tilde{a}_{p}$ are expanded to different orders. 
\cref{fig:ADC_matrices} shows the perturbative structure of these matrices for the ADC approximations employed in this work (ADC(2), ADC(2)-X, and ADC(3)).
Explicit expressions for $\tilde{H}^{(l)}$ and $\tilde{a}_{p}^{(l)}$ are obtained from the Baker--Campbell--Hausdorff (BCH) expansions of $\tilde{{H}}$ and $\tilde{a}_{p}$, e.g.:
\begin{align}
\label{eqn:bch_h}
\tilde{H} &= H^{(0)} + V + [H^{(0)},T^{(1)} - T^{(1){\dagger}}] + [H^{(0)},T^{(2)} - T^{(2){\dagger}}]\notag   \\
&+\frac{1}{2!}[V+(V + [H^{(0)},T^{(1)} - T^{(1){\dagger}}]),T^{(1)}-T^{(1){\dagger}}]+...
\end{align}
These equations depend on the amplitudes of excitation operators $T^{(k)}$ 
 \begin{align}
	\label{eqn:ex_op}
	T^{(k)} &= \sum_m^N T^{(k)}_{m}, \notag \\
	T^{(k)}_{m} &= \frac{1}{(m!)^{2}}\sum_{ijab\ldots} t_{ij\ldots}^{ab\ldots(k)}a_{a}^{\dagger}a_{b}^{\dagger}\ldots a_{j}a_{i}
\end{align}
that are calculated  by solving a system of projected amplitude equations.\cite{banerjee:2019p224112}
The low-order ADC approximations (ADC($n$), $n$ $\le$ 3) require calculating up to the $n$th-order single-excitation amplitudes ($t_{i}^{a(n)}$) and up to the ($n-1$)th order double-excitation amplitudes ($t_{ij}^{ab(n-1)}$).
Additionally, the third-order contributions to the $\bold{T}_{\pm}$ matrix of ADC(3) formally depend on the triple-excitation amplitudes ($t_{ijk}^{abc(k)}$).\cite{Trofimov:1999p9982,Hodecker:2022p074104}
In practice, the $t_{ijk}^{abc(k)}$ contributions are very small and we neglect them in our implementation of ADC(3).   
   
\subsection{Quantifying the spin contamination in ADC calculations}
\label{sec:theory:spin_contamination}

The focus of this work is to investigate the spin contamination in single-reference ADC calculations of charged excited states (EA/IP-ADC) for open-shell molecules.
The accuracy of ADC approximations strongly depends on the quality of underlying Hartree--Fock reference wavefunction.
In particular for open-shell molecules, molecular orbitals and orbital energies appearing in the zeroth-order ADC  Hamiltonian in \cref{eqn:h_zero} are usually computed using the unrestricted Hartree--Fock theory (UHF), which may introduce spin contamination into the ADC excited-state energies and properties.
Spin contamination can be calculated by subtracting the computed expectation value of the spin-squared operator ($S^2$) from its exact eigenvalue for a particular electronic state:\cite{stuck:2013p244109,andrews:1991p423,krylov:2017p151} 
\begin{align}
	\label{eqn:spin_contamination}
	\Delta\mathit{S}^{2}\equiv \braket{S^{2}}_{computed}-\braket{S^{2}}_{exact} \ .
\end{align}
Here, we present an approach based on effective Liouvillian theory that allows to compute the $S^2$ expectation values and spin contamination in the ADC calculations.
To the best of our knowledge, this work is the first study of excited-state spin contamination in ADC. 
An investigation of spin contamination in the related CC2 method has been reported recently.\cite{kitsaras:2021p131101}

A general expression for the expectation value of spin-squared operator with respect to a wavefunction $\ket{\Psi}$ has the form:\cite{kouba:1969p513,purvis:1988p2203,Stanton:1994p371,krylov:2000p6052} 
\begin{align}
\label{eqn:spin_squared}
\braket{S^{2}} &=  \frac{1}{4}(\sum_{p\in \alpha} \gamma_p^p - \sum_{\bar{p}\in \beta} \gamma_{\bar{p}}^{\bar{p}})^{2} \notag \\
 & + \frac{1}{2}(\sum_{p\in \alpha} \gamma_p^p + \sum_{\bar{p}\in \beta} \gamma_{\bar{p}}^{\bar{p}}) 
+ \sum_{\substack{p,s\in\alpha \\ \bar{q},\bar{r}\in\beta}}S^{p}_{\bar{r}}S^{\bar{q}}_{s}\Gamma^{p\bar{q}}_{\bar{r}s} \, ,
\end{align}
where we use a bar to distinguish between the spin-orbitals with $\alpha$ and $\beta$ spin, $\gamma^p_q$ and $\Gamma^{pq}_{rs}$ are the one- and two-particle reduced density matrices
\begin{align}
	\label{eqn:rdms}
	\gamma^p_q &= \braket{\Psi|\c{p}\a{q}|\Psi} \, , \quad
	\Gamma^{pq}_{rs} = \braket{\Psi|\c{p}\c{q}\a{s}\a{r}|\Psi} \, ,
\end{align}
and $S^p_{\bar{q}}$ is the overlap of two spatial molecular orbitals with opposite spin:
\begin{align}
	\label{eqn:mo_overlap}
	S^p_{\bar{q}} = \braket{\phi_p | \phi_q} \, ,\quad \phi_p \in \alpha, \ \phi_q \in \beta .
\end{align}
Evaluating $\braket{S^{2}}$ for a charged excited state $\ket{\Psi}$ requires calculating the excited-state one- and two-particle reduced density matrices (\cref{eqn:rdms}), which are the expectation values of one- and two-body operators ($O = \c{p}\a{q},\ \c{p}\c{q}\a{s}\a{r}$) with respect to $\ket{\Psi}$. 
An approach to evaluate the operator expectation values in ADC using intermediate state representation has been presented by Trofimov, Dempwolff, and co-workers.\cite{Schirmer:2004p11449,Trofimov:2005p144115,Knippenberg:2012p64107,plasser:2014p24106,plasser2014p24107,dempwolff:2020p024113,dempwolff:2020p024125,Dempwolff:2021p104117}
Below, we demonstrate how the ADC excited-state reduced density matrices and the expectation values of spin-squared operator can be computed within the framework of effective Liouvillian theory. 

The expectation value of an arbitrary operator $O$ for an excited state described by the ADC eigenvector $Y_{\pm\mu I}$ has the form:
\begin{align}
	\label{eqn:op_exp_value_adc}
	\braket{O}_{\pm I} = \sum_{(n)}^{(m)} \sum_{\mu \nu} Y^{\dag}_{\pm I \mu} O^{(n)}_{\pm\mu\nu} Y_{\pm \nu I} \ ,
\end{align}
where $O^{(n)}_{\pm\mu\nu}$ is the $n$th-order matrix element of the operator $O$ and the summation over ($n$) includes all contributions up to the order of ADC approximation ($m$).
In order to obtain the expressions for $O^{(n)}_{\pm\mu\nu}$, we separate these matrix elements into a reference (ground-state) contribution $O^{(n)}_{0}$ and the excitation component $\Delta O^{(n)}_{\pm\mu\nu}$
\begin{align}
	\label{eqn:op_exp_value_adc_separated}
	O^{(n)}_{\pm\mu\nu} = O^{(n)}_{0}\delta_{\mu\nu} \pm \Delta O^{(n)}_{\pm\mu\nu} \ .
\end{align}
Equations for $\Delta O^{(n)}_{\pm\mu\nu}$ can be obtained by analogy with \cref{eqn:M_plus,eqn:M_minus} for the matrix elements of the shifted effective Hamiltonian.
Simplifying \cref{eqn:M_plus,eqn:M_minus} and replacing $H$ by $O$, we obtain:
\begin{align}
	\label{eqn:dO_plus}
	\Delta O_{+\mu\nu}^{(n)} = \sum_{klm}^{k+l+m=n} &\left( \braket{\Phi|h_{+\mu}^{(k)}\tilde{O}^{(l)}h_{+\nu}^{(m)\dagger}|\Phi} \right. \notag \\
	&\left. - \delta_{\mu\nu} \delta_{km}\braket{\Phi|\tilde{O}^{(l)}|\Phi} \right) \ , \\
	\label{eqn:dO_minus}
	\Delta O_{-\mu\nu}^{(n)} = \sum_{klm}^{k+l+m=n} &\left( -\braket{\Phi|h_{-\nu}^{(m)} \tilde{O}^{(l)} h_{-\mu}^{(k)\dagger}|\Phi} \right. \notag \\
	&\left.+ \delta_{\mu\nu} \delta_{km}\braket{\Phi|\tilde{O}^{(l)}|\Phi}\right) \ , 
\end{align}
where $\tilde{O}^{(l)}$ is the $l$th-order contribution to an effective operator $\tilde{O}=e^{-(T-T^{\dagger})}Oe^{(T-T^{\dagger})}$ that can be obtained from its BCH expansion
\begin{align}
	\label{eqn:gammas_bch}
	\tilde{O} &= \mathit{O}^{(0)} + [\mathit{O}^{(0)},T^{(1)}-T^{(1){\dagger}}]+ [\mathit{O}^{(0)},T^{(2)}-T^{(2){\dagger}}]\notag\\
	&+\frac{1}{2!}[[\mathit{O}^{(0)},T^{(1)}-T^{(1){\dagger}}],T^{(1)}-T^{(1){\dagger}}]+\dots 
\end{align}
Identifying the second term on the r.h.s.\@ of \cref{eqn:dO_plus,eqn:dO_minus} as $O^{(n)}_{0}\delta_{\mu\nu}$, moving it to the l.h.s., and inverting the sign of \cref{eqn:dO_minus}, we obtain the expressions for $O^{(k)}_{\pm\mu\nu}$ of an arbitrary operator:
\begin{align}
	\label{eqn:O_plus}
	O^{(n)}_{+\mu\nu} &= \sum_{klm}^{k+l+m=n} \braket{\Phi|h_{+\mu}^{(k)}\tilde{O}^{(l)}h_{+\nu}^{(m)\dagger}|\Phi} \ , \\
	\label{eqn:O_minus}
	O_{-\mu\nu}^{(n)}  &= \sum_{klm}^{k+l+m=n} \braket{\Phi|h_{-\nu}^{(m)} \tilde{O}^{(l)} h_{-\mu}^{(k)\dagger}|\Phi} \ .
\end{align}
Combining \cref{eqn:op_exp_value_adc} with \cref{eqn:O_plus,eqn:O_minus} for $O = \c{p}\a{q} $ and $ \c{p}\c{q}\a{s}\a{r}$, we obtain the working equations for one- and two-particle reduced density matrices of charged excited states for an arbitrary-order ADC approximation.
As an example, we present the equations for $\gamma^p_q$ and $\Gamma^{pq}_{rs}$ of ADC(2) in the Supplementary Information.
Once the excited-state $\gamma^p_q$ and $\Gamma^{pq}_{rs}$ are computed, we evaluate the corresponding expectation values of spin-squared operator and spin contamination according to \cref{eqn:spin_contamination,eqn:spin_squared}.

\subsection{Reducing the spin contamination in ADC using the ROHF and OMP($\mathbf{n}$) reference orbitals}
\label{sec:theory:sr_adc_reference_wavefunctions}

To reduce the spin contamination in ADC calculations of charged excitations for open-shell molecules, we combined our EA/IP-ADC implementation with the reference orbitals from i) restricted open-shell Hartree--Fock (ROHF)\cite{roothaan:1960p179} and ii) orbital-optimized $n$th-order M\o ller--Plesset perturbation (OMP($n$)) theories.\cite{bozkaya:2013p184103,bozkaya:2011p224103,bozkaya:2011p104103,Lee:2019p244106,bertels:2019p4170,Neese:2009p3060,lochan:2007p164101,kurlancheek:2009p1223}
These reference wavefunctions are known to either completely eradicate or significantly reduce the spin contamination in state-specific (usually, ground-state) calculations of open-shell systems.\cite{knowles:1991p130,lochan:2007p164101,Neese:2009p3060,bozkaya:2014p4389,soydas:2015p1564,kitsaras:2021p131101}

To combine EA/IP-ADC with ROHF, we use the approach developed by Knowles and co-workers\cite{knowles:1991p130} in restricted open-shell M\o ller--Plesset perturbation theory. 
Following this approach, the ROHF-based ADC methods (ADC($n$)/ROHF) can be implemented by modifying the unrestricted ADC implementation (ADC($n$)/UHF) as described below: 
\begin{enumerate}
	\item Using the ROHF orbitals we calculate the Fock matrices (\cref{eqn:fock_matrix}) for spin-up ($\alpha$) and spin-down ($\beta$) electrons and semicanonicalize them following the procedure described in Ref.\@ \citenum{knowles:1991p130}.
	The diagonal Fock matrix elements calculated in the semicanonical basis are used to define the ADC zeroth-order Hamiltonian in \cref{eqn:h_zero}.
	The off-diagonal Fock matrix elements ($\f{i}{a}$) are included in the perturbation operator $V$, which now has the form:
	\begin{align}
		\label{eqn:perturbation_rohf}
		V &= \frac{1}{4} \sum_{pqrs} \v{pq}{rs} \{ \c{p}\c{q}\a{s}\a{r} \} \notag \\
		&+ \sum_{ia} \left(\f{i}{a} \{ \c{i} \a{a} \} 
		+ \f{a}{i} \{ \c{a} \a{i}\} \right) 
		\ ,
	\end{align}
	\item The $\f{i}{a}$ terms in $V$ significantly modify the ADC equations. 
	First, these contributions give rise to the $\f{i}{a}$-dependent terms in effective Hamiltonian matrix $\bold{M}_{\pm}$.
	Second, for $V$ in \cref{eqn:perturbation_rohf} the Brillouin's theorem is no longer satisfied and the first-order single-excitation amplitudes in excitation operator $T^{(1)}$ (\cref{eqn:ex_op}) have non-zero values:
	\begin{align}
		\label{eqn:first_order_singles}
		\t{i}{a(1)} = \frac{\f{a}{i}}{\e{i}-\e{a}} \ .
	\end{align}
	The $\t{i}{a(1)}$ terms enter the projected amplitude equations for other amplitudes and give rise to new contributions in the equations for all ADC matrices.
\end{enumerate}

An alternative approach to mitigate spin contamination that we explore in this work is to combine the ADC($n$) methods with orbitals from the OMP($n$) calculations (ADC($n$)/OMP($n$)).
Orbital optimization has been shown to significantly reduce spin contamination and improve the accuracy of correlated (post-Hartree--Fock) calculations for open-shell molecules.\cite{scuseria:1987p354,krylov:1998p10669,Sherrill:1998p4171,gwaltney:2000p3548,lochan:2007p164101,bozkaya:2011p224103,bozkaya:2011p104103,bozkaya:2012p204114,bozkaya:2013p184103,bozkaya:2013p104116,Bozkaya:2013p54104,Sokolov:2013p024107}
Considering the close relationship between ADC and M\o ller--Plesset perturbation theories, using the OMP($n$) orbitals to perform the ADC($n$) calculations is a promising alternative to the UHF reference orbitals for molecules with unpaired electrons. 

We refer the readers interested in details of orbital-optimized M\o ller--Plesset perturbation theory to excellent publications on this subject.\cite{bozkaya:2011p224103,bozkaya:2011p104103,lochan:2007p164101}
As for the ROHF reference, combining ADC($n$) with the OMP($n$) orbitals requires several modifications to the UHF-based ADC methods as outlined below:
\begin{enumerate}
	\item Upon successful completion of the reference OMP($n$) calculation, the Fock matrices for $\alpha$- and $\beta$-electrons are semicanonicalized as described for ADC($n$)/ROHF. 
	The diagonal Fock matrix elements are used to define the ADC zeroth-order Hamiltonian in \cref{eqn:h_zero}, while the off-diagonal $\f{i}{a}$ elements enter the expression for the perturbation operator in \cref{eqn:perturbation_rohf}.
	\item The $\f{i}{a}$ contributions in \cref{eqn:perturbation_rohf} give rise to new terms in the effective Hamiltonian matrix $\bold{M}_{\pm}$.
	\item As in the reference OMP($n$) calculations, all single-excitation amplitudes and the corresponding projections of the effective Hamiltonian are assumed to be zero in all ADC equations ($\t{i}{a(k)}\approx 0$ and $\braket{\Phi|a_{i}^{\dagger}a_{a}\tilde{H}^{(k)}|\Phi} \approx 0$, $\forall k$).	
	\item The converged OMP($n$) double-excitation amplitudes ($\t{ij}{ab(k)}$) are transformed to the semicanonical basis\cite{bozkaya:2012p204114} and are used to compute the ADC matrix elements.
\end{enumerate}
Working equations for the ROHF-based EA/IP-ADC($n$) methods are provided in the Supplementary Information. 
The equations for EA/IP-ADC($n$) combined with the OMP($n$) reference orbitals can be obtained by setting the terms depending on single-excitation amplitudes to zero.

\section{Computational details}
\label{sec:comp_details}

The ROHF- and OMP($n$)-based EA/IP-ADC(2), EA/IP-ADC(2)-X, and EA/IP-ADC(3) methods were implemented in the developer's version of PySCF\cite{sun:2020p24109} by modifying the existing unrestricted ADC module. 
Additionally, for each reference (UHF, ROHF, and OMP($n$)) we implemented the subroutines for calculating one- and two-particle reduced density matrices (\cref{eqn:rdms}), the expectation values of spin-squared operator (\cref{eqn:spin_squared}), and spin contamination (\cref{eqn:spin_contamination}). 
Our implementation of the ADC reduced density matrices was validated by reproducing the EA/IP-ADC(2)/UHF dipole moments of charged states reported by Dempwolff and co-workers.\cite{dempwolff:2020p024113,Dempwolff:2021p104117}
For EA/IP-ADC(3), the reduced density matrices incorporated contributions up to the third order in perturbation theory.
We are aware of only one previous implementation of EA/IP-ADC($n$) with the ROHF reference in the Q-Chem package.\cite{epifanovsky:2021p84801}
Numerical tests of both EA/IP-ADC($n$)/ROHF programs (PySCF and Q-Chem) indicate that they produce very similar results when the contributions from first-order single excitations (\cref{eqn:first_order_singles}) are neglected in our PySCF implementation, suggesting that these excitations may be missing in the Q-Chem implementation of ADC($n$)/ROHF.
To the best of our knowledge, our work represents the first implementation of the EA/IP-ADC($n$)/OMP($n$) methods.
Specifically, the EA/IP-ADC(2) and EA/IP-ADC(2)-X methods were combined with the OMP(2) reference orbitals, while for EA/IP-ADC(3) we used the OMP(3) reference.

\begin{table}[t!]
	\captionsetup{justification=raggedright,singlelinecheck=false,font=footnotesize}
	\caption{Open-shell molecules studied in this work and their ground-state spin contamination (a.u.) computed using UHF, MP($n$) ($n$ = 2, 3) with the UHF reference, orbital-optimized MP($n$) (OMP($n$)), and MP($n$) with the ROHF reference. }
	\label{tab:ref_sc}
	\setlength{\extrarowheight}{2pt}
	\setstretch{1}
	\tiny
	\centering
	\hspace*{-0.8cm}
	\begin{threeparttable}
		\begin{tabular}{llccccccc}
			\hline\hline
			System          & State                  &   UHF   & MP(2)  & MP(3)  & OMP(2)  & OMP(3) & MP(2)  & MP(3)  \\
			                &                        &         &  UHF   &  UHF   &         &        &  ROHF  &  ROHF  \\ \hline
			\ce{BeH   }     & ${}^{2}\Sigma^{+}    $ & $0.00 $ & $0.00$ & $0.00$ & $0.00$  & $0.00$ & $0.00$ & $0.00$ \\
			\ce{OH    }     & ${}^{2}\Pi           $ & $0.01 $ & $0.00$ & $0.00$ & $0.00$  & $0.00$ & $0.00$ & $0.01$ \\
			\ce{NH2   }     & ${}^{2}B_{1}         $ & $0.01 $ & $0.00$ & $0.00$ & $0.00$  & $0.00$ & $0.00$ & $0.01$ \\
			\ce{SH    }     & ${}^{2}\Pi           $ & $0.01 $ & $0.00$ & $0.00$ & $0.00$  & $0.00$ & $0.01$ & $0.01$ \\
			\ce{CH3   }     & ${}^{2}A^{''}_{1}    $ & $0.01 $ & $0.00$ & $0.00$ & $0.00$  & $0.00$ & $0.00$ & $0.01$ \\
			\ce{SF    }     & ${}^{2}\Pi           $ & $0.01 $ & $0.00$ & $0.00$ & $0.00$  & $0.00$ & $0.01$ & $0.01$ \\
			\ce{OOH   }     & ${}^{2}A''           $ & $0.01 $ & $0.00$ & $0.00$ & $0.00$  & $0.00$ & $0.01$ & $0.01$ \\
			\ce{CH2   }     & ${}^{3}B_{1}         $ & $0.02 $ & $0.00$ & $0.00$ & $0.00$  & $0.00$ & $0.01$ & $0.01$ \\
			\ce{NH    }     & ${}^{3}\Sigma^{-}    $ & $0.02 $ & $0.00$ & $0.00$ & $0.00$  & $0.00$ & $0.01$ & $0.01$ \\
			\ce{PH2   }     & ${}^{2}B_{1}         $ & $0.02 $ & $0.00$ & $0.00$ & $0.00$  & $0.00$ & $0.01$ & $0.01$ \\
			\ce{Si2   }     & ${}^{3}\Sigma^{-}    $ & $0.02 $ & $0.00$ & $0.00$ & $0.00$  & $0.00$ & $0.01$ & $0.01$ \\
			\ce{SiF   }     & ${}^{2}\Pi           $ & $0.02 $ & $0.00$ & $0.01$ & $0.00$  & $0.00$ & $0.01$ & $0.01$ \\
			\ce{FO    }     & ${}^{2}\Pi           $ & $0.04 $ & $0.02$ & $0.02$ & $0.00$  & $0.00$ & $0.00$ & $0.01$ \\
			\ce{O2    }     & ${}^{3}\Sigma^{-}_{g}$ & $0.05 $ & $0.01$ & $0.02$ & $0.00$  & $0.00$ & $0.02$ & $0.02$ \\
			\ce{S2    }     & ${}^{3}\Sigma^{-}_{g}$ & $0.05 $ & $0.01$ & $0.02$ & $0.00$  & $0.00$ & $0.02$ & $0.02$ \\
			\ce{BO    }     & ${}^{2}\Sigma^{+}    $ & $0.06 $ & $0.03$ & $0.04$ & $0.00$  & $0.01$ & $0.00$ & $0.01$ \\
			\ce{BN    }     & ${}^{3}\Pi           $ & $0.06 $ & $0.04$ & $0.04$ & $0.00$  & $0.00$ & $0.00$ & $0.01$ \\
			\ce{SO    }     & ${}^{3}\Sigma^{-}    $ & $0.06 $ & $0.02$ & $0.02$ & $0.00$  & $0.00$ & $0.02$ & $0.02$ \\
			\ce{NO    }     & ${}^{2}\Pi           $ & $0.10 $ & $0.07$ & $0.07$ & $0.00$  & $0.00$ & $0.01$ & $0.01$ \\
			\ce{NCO   }     & ${}^{2}\Pi           $ & $0.11 $ & $0.06$ & $0.06$ & $0.00$  & $0.01$ & $0.01$ & $0.01$ \\
			\ce{AlO   }     & ${}^{2}\Sigma^{+}    $ & $0.13 $ & $0.09$ & $0.10$ & $0.00$  & $0.00$ & $0.01$ & $0.01$ \\
			\ce{CNC   }     & ${}^{2}\Pi_{g}       $ & $0.14 $ & $0.07$ & $0.08$ & $0.00$  & $0.00$ & $0.02$ & $0.03$ \\
			\ce{NO2   }     & ${}^{2}A_{1}         $ & $0.14 $ & $0.10$ & $0.11$ & $0.00$  & $0.00$ & $0.01$ & $0.01$ \\
			\ce{CH2CHO}     & ${}^{2}A''           $ & $0.19 $ & $0.11$ & $0.12$ & $0.00$  & $0.01$ & $0.01$ & $0.01$ \\
			\ce{C4O    }    & ${}^{3}\Sigma^{-}    $ & $0.21 $ & $0.12$ & $0.13$ & $0.02$  & $0.02$ & $0.03$ & $0.02$ \\
			\ce{BP}         & ${}^{3}\Pi           $ & $0.22 $ & $0.17$ & $0.18$ & $0.00$  & $0.00$ & $0.01$ & $0.01$ \\
			\ce{C3H5    }   & ${}^{2}A_{2}         $ & $0.22 $ & $0.13$ & $0.13$ & $-0.01$ & $0.01$ & $0.03$ & $0.04$ \\
			\ce{N3    }     & ${}^{2}\Pi_{g}       $ & $0.23 $ & $0.14$ & $0.15$ & $0.01$  & $0.01$ & $0.03$ & $0.02$ \\
			\ce{SCN   }     & ${}^{2}\Pi           $ & $0.24 $ & $0.14$ & $0.15$ & $0.00$  & $0.01$ & $0.01$ & $0.01$ \\
			\ce{CH2CN }     & ${}^{2}B_{1}         $ & $0.24 $ & $0.14$ & $0.15$ & $0.00$  & $0.01$ & $0.01$ & $0.01$ \\
			\ce{C2H3  }     & ${}^{2}A'            $ & $0.25 $ & $0.17$ & $0.18$ & $0.00$  & $0.01$ & $0.00$ & $0.01$ \\
			\ce{CNN   }     & ${}^{3}\Sigma^{-}    $ & $0.26 $ & $0.14$ & $0.14$ & $0.00$  & $0.01$ & $0.03$ & $0.04$ \\
			\ce{C3H3   }    & ${}^{2}B_{1}         $ & $0.27 $ & $0.16$ & $0.17$ & $0.00$  & $0.01$ & $0.01$ & $0.02$ \\
			\ce{CH    }     & ${}^{2}\Pi           $ & $0.33 $ & $0.33$ & $0.33$ & $0.00$  & $0.00$ & $0.01$ & $0.01$ \\
			\ce{CHN2  }     & ${}^{2}A''           $ & $0.36 $ & $0.24$ & $0.25$ & $0.00$  & $0.01$ & $0.02$ & $0.02$ \\
			\ce{HCCN  }     & ${}^{3}A''           $ & $0.40 $ & $0.23$ & $0.24$ & $-0.01$ & $0.02$ & $0.02$ & $0.03$ \\
			\ce{CN    }     & ${}^{2}\Sigma^{+}    $ & $0.49 $ & $0.36$ & $0.36$ & $0.00$  & $0.01$ & $0.01$ & $0.01$ \\
			\ce{NS    }     & ${}^{2}\Pi           $ & $0.51 $ & $0.41$ & $0.42$ & $0.00$  & $0.01$ & $0.01$ & $0.01$ \\
			\ce{SiC   }     & ${}^{3}\Pi           $ & $0.58 $ & $0.49$ & $0.49$ & $0.00$  & $0.02$ & $0.01$ & $0.02$ \\
			\ce{CP    }     & ${}^{2}\Sigma^{+}    $ & $0.97 $ & $0.79$ & $0.79$ & $0.00$  & $0.05$ & $0.01$ & $0.01$ \\ \hline
			Average &                        & $0.18$  & $0.12$ & $0.12$ & $0.00$  & $0.01$ & $0.01$ & $0.01$ \\
			Sum     &                        & $7.03$  & $4.84$ & $4.98$ & $0.02$  & $0.25$ & $0.45$ & $0.54$ \\ \hline\hline
		\end{tabular}
	\end{threeparttable}
\end{table}   

To benchmark the accuracy of EA/IP-ADC($n$) with the UHF, ROHF, and OMP($n$) references, we performed the calculations of vertical electron affinities (EA), vertical ionization energies (IP), and excited-state spin contamination for the lowest-energy charged states of 40 neutral open-shell molecules shown in \cref{tab:ref_sc}. 
These systems were classified into two groups based on the spin contamination (\SC) in reference UHF calculation: (i) 18 weakly spin-contaminated molecules with \SC $< 0.1$ a.u.\@ and (ii) 22 strongly spin-contaminated molecules with \SC $ \ge 0.1$ a.u.
The equilibrium geometries of all molecules were computed using coupled cluster theory with single, double, and perturbative triple excitations combined with the ROHF reference (CCSD(T)/ROHF)\cite{watts:1993p8718,Raghavachari:1989p479,Bartlett:1990p513,Crawford:2000p33,Shavitt:2009} and are reported in the Supplementary Information. 
The vertical EA's and IP's from CCSD(T)/ROHF were used to benchmark the accuracy of EA/IP-ADC($n$) with the UHF, ROHF, and OMP($n$) references.
For comparison, we also report the EA's and IP's from equation-of-motion coupled cluster theory with single and double excitations (EOM-CCSD).\cite{Nooijen:1992p55,stanton:1993p7029,Nooijen:1993p15,Nooijen:1995p1681,Crawford:2000p33,Krylov:2008p433,Shavitt:2009}
In all calculations, the aug-cc-pVDZ basis set\cite{dunning:1989p1007} was used. 
All coupled cluster results were obtained using the Q-Chem\cite{shao:2015p184} and CFOUR\cite{matthews:2020p214108} software packages.
Throughout the manuscript, positive electron affinity indicates exothermic electron attachment (i.e., EA = $E_{N}$ $-$ $E_{N+1}$) while a positive ionization energy corresponds to an endothermic process (IP = $E_{N-1}$ $-$ $E_{N}$). 

\section{Results and Discussion}
\label{sec:results}
\subsection{
	Benchmarking EA/IP-ADC with the unrestricted (UHF) reference
}
\label{sec:results:UHF}

\begin{figure}[t!]
	\centering
	\captionsetup{justification=raggedright,singlelinecheck=false,font=footnotesize}
	\includegraphics[scale=0.30,trim=0.0cm 0.0cm 0.0cm 0.0cm,clip]{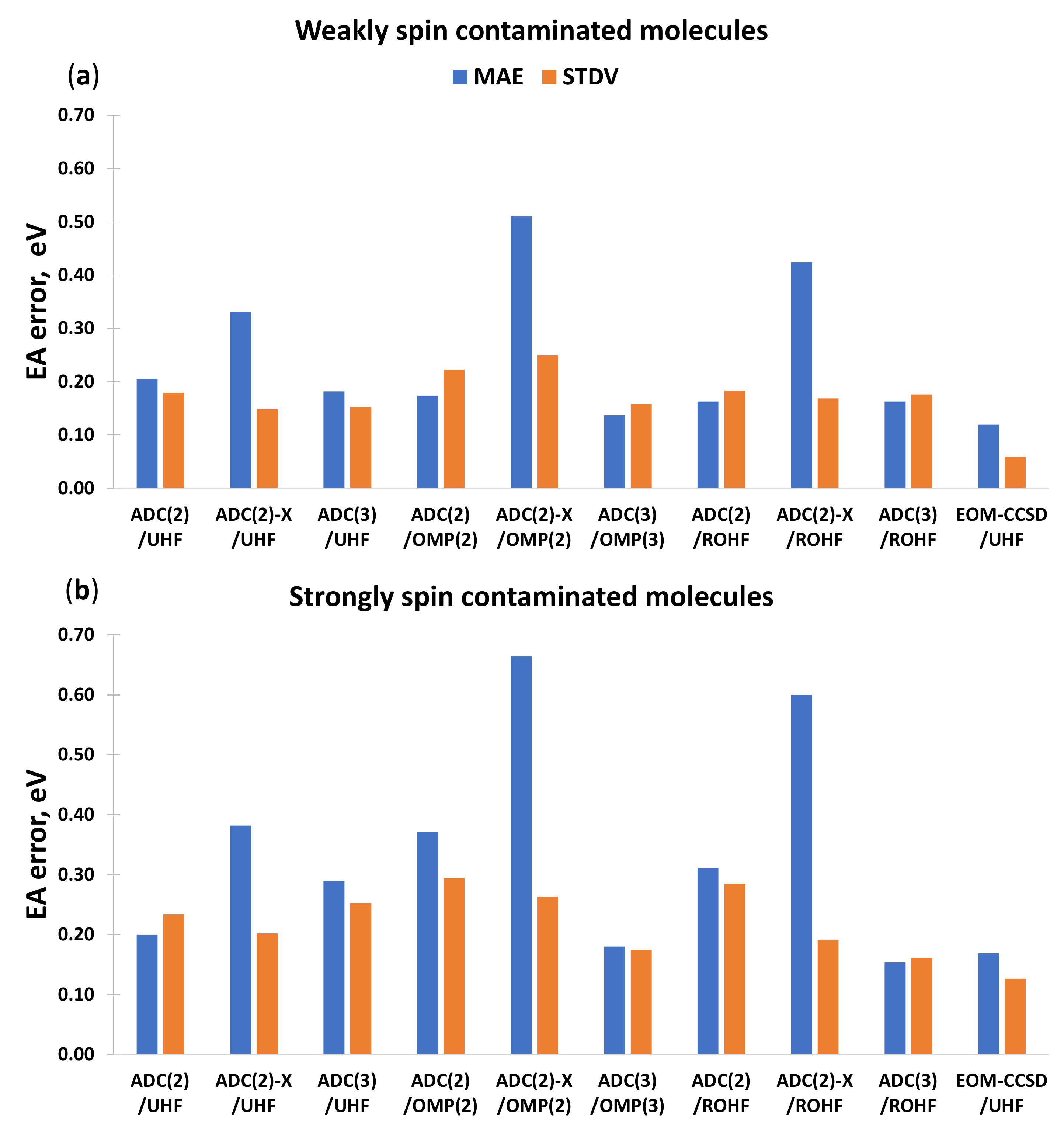}
	\caption{Mean absolute errors (MAE) and standard deviations (STDV) in the EA-ADC vertical electron affinities (eV) for (a) 18 weakly and (b) 22 strongly spin contaminated molecules with the ground-state UHF spin contamination of $< 0.1$ and $\ge 0.1$ a.u., respectively. 
		Reference data is from CCSD(T).
		The aug-cc-pVDZ basis set was used. 
		See \cref{tab:ADC_EA_weak,tab:ADC_EA_strong} for data on individual molecules.
	}
	\label{fig:ADC_EA}
\end{figure}

We begin by analyzing the errors in vertical electron affinities (EA's) computed using EA-ADC/UHF for weakly and strongly spin contaminated open-shell molecules (WSM and SSM), as defined in \cref{sec:comp_details}.
The EA-ADC/UHF and EA-EOM-CCSD/UHF electron affinities for both sets of molecules are shown in  \cref{tab:ADC_EA_weak,tab:ADC_EA_strong}, along with the reference results from CCSD(T)/ROHF. 
\cref{fig:ADC_EA} illustrates how the mean absolute errors (\mae) and standard deviations of errors (\std) change with the increasing level of EA-ADC theory. 
For WSM, \mae reduces in the following order: EA-ADC(2)-X (0.33 eV) $>$ EA-ADC(2) (0.21 eV) $>$ EA-ADC(3) (0.18 eV), in a very good agreement with the benchmark results from Ref.\@ \citenum{banerjee:2019p224112}.
This trend changes for SSM where the \mae of EA-ADC(2) remains low (0.20 eV) but EA-ADC(2)-X and EA-ADC(3) show significantly larger errors than for WSM (\mae = 0.38 and 0.29 eV, respectively).
Increasing spin contamination in the UHF reference wavefunction also affects \std for all three methods, which grows by at least 33\% for EA-ADC(2) and as much as 80\% for EA-ADC(3).
The EA-EOM-CCSD/UHF method shows the smallest \mae and \std out of all UHF-based methods considered in this study for SSM, indicating that it is less sensitive to significant spin contamination in the UHF reference than EA-ADC.

\begin{table*}[t!]
	\captionsetup{justification=raggedright,singlelinecheck=false,font=footnotesize}
	\caption{
		Vertical electron affinities (eV) of weakly spin contaminated open-shell molecules computed using EA-ADC, EA-EOM-CCSD, and CCSD(T) with various reference wavefunctions (UHF, OMP, and ROHF). 
		The aug-cc-pVDZ basis set was used in all calculations. 
		Also shown are mean absolute errors (\mae) and standard deviations (\std) relative to the CCSD(T) results. 
	}
	\label{tab:ADC_EA_weak}
	\setlength{\extrarowheight}{2pt}
	\setstretch{1}
	\tiny
	\centering
	\hspace*{-0.8cm}
	\begin{threeparttable}
		\begin{tabular}{llccccccccccc}
			\hline\hline
			System      & Transition                                      & ADC(2)  & ADC(2)-X & ADC(3)  & ADC(2)  & ADC(2)-X & ADC(3)  & ADC(2)  & ADC(2)-X & ADC(3)  & EOM-CCSD & CCSD(T) \\
			            &                                                 &   UHF   &   UHF    &   UHF   & OMP(2)  &  OMP(2)  & OMP(3)  &  ROHF   &   ROHF   &  ROHF   &   UHF    &  ROHF   \\ \hline
			\ce{BeH   } & ${}^{2}\Sigma^{+}\rightarrow {}^{1}\Sigma^{+}$  & $-0.01$ & $ 0.53$  & $ 0.47$ & $ 0.00$ & $ 0.54$  & $ 0.48$ & $ 0.00$ & $ 0.54$  & $ 0.48$ &  $0.42$  & $0.48$  \\
			\ce{OH    } & ${}^{2}\Pi \rightarrow {}^{1}\Sigma^{+}$        & $ 1.33$ & $ 2.11$  & $ 1.22$ & $ 1.80$ & $ 2.60$  & $ 1.21$ & $ 1.41$ & $ 2.22$  & $ 1.27$ &  $1.50$  & $1.63$  \\
			\ce{NH2   } & ${}^{2}B_{1}\rightarrow {}^{1}A_{1}$            & $ 0.35$ & $ 0.95$  & $ 0.16$ & $ 0.64$ & $ 1.26$  & $ 0.22$ & $ 0.44$ & $ 1.07$  & $ 0.22$ &  $0.41$  & $0.54$  \\
			\ce{SH    } & ${}^{2}\Pi\rightarrow {}^{1}\Sigma^{+}$         & $ 1.99$ & $ 2.47$  & $ 2.01$ & $ 2.10$ & $ 2.60$  & $ 2.04$ & $ 2.05$ & $ 2.55$  & $ 2.05$ &  $2.05$  & $2.12$  \\
			\ce{CH3   } & ${}^{2}A^{''}_{1}\rightarrow {}^{1}A^{'}_{1}$   & $-0.49$ & $ 0.03$  & $-0.47$ & $-0.37$ & $ 0.18$  & $-0.39$ & $-0.40$ & $ 0.15$  & $-0.40$ & -$0.36$  & $-0.24$ \\
			\ce{SF    } & ${}^{2}\Pi\rightarrow {}^{1}\Sigma^{+}$         & $ 1.98$ & $ 2.45$  & $ 2.00$ & $ 2.04$ & $ 2.51$  & $ 2.08$ & $ 2.05$ & $ 2.53$  & $ 2.04$ &  $2.02$  & $2.08$  \\
			\ce{OOH   } & ${}^{2}A''\rightarrow {}^{1}A'$                 & $ 0.41$ & $ 1.00$  & $ 0.31$ & $ 0.68$ & $ 1.31$  & $ 0.31$ & $ 0.56$ & $ 1.17$  & $ 0.38$ &  $0.24$  & $0.47$  \\
			\ce{CH2   } & ${}^{3}B_{1}\rightarrow {}^{2}B_{1}$            & $-0.42$ & $ 0.23$  & $-0.19$ & $-0.30$ & $ 0.38$  & $-0.11$ & $-0.36$ & $ 0.32$  & $-0.14$ & -$0.11$  & $0.02$  \\
			\ce{NH    } & ${}^{3}\Sigma^{-}\rightarrow {}^{2}\Pi  $       & $-0.36$ & $ 0.37$  & $-0.18$ & $-0.14$ & $ 0.63$  & $-0.10$ & $-0.27$ & $ 0.50$  & $-0.11$ & -$0.06$  & $0.06$  \\
			\ce{PH2   } & ${}^{2}B_{1}\rightarrow {}^{1}A_{1}$            & $ 0.89$ & $ 1.32$  & $ 0.95$ & $ 0.98$ & $ 1.41$  & $ 1.00$ & $ 0.97$ & $ 1.42$  & $ 1.01$ &  $0.97$  & $1.04$  \\
			\ce{Si2   } & ${}^{3}\Sigma^{-}\rightarrow {}^{2}\Pi_{u}  $   & $ 2.01$ & $ 2.35$  & $ 1.85$ & $ 2.06$ & $ 2.39$  & $ 1.84$ & $ 2.04$ & $ 2.38$  & $ 1.87$ &  $1.85$  & $1.94$  \\
			\ce{SiF   } & ${}^{2}\Pi \rightarrow {}^{3}\Sigma^{-}$        & $ 0.70$ & $ 1.02$  & $ 0.84$ & $ 0.69$ & $ 1.02$  & $ 0.86$ & $ 0.74$ & $ 1.07$  & $ 0.87$ &  $0.71$  & $0.78$  \\
			\ce{FO    } & ${}^{2}\Pi\rightarrow {}^{1}\Sigma_{g}^{+}$     & $ 1.91$ & $ 2.46$  & $ 1.47$ & $ 2.06$ & $ 2.69$  & $ 1.62$ & $ 2.10$ & $ 2.62$  & $ 1.50$ &  $1.66$  & $1.83$  \\
			\ce{O2    } & ${}^{3}\Sigma^{-}_{g}\rightarrow {}^{2}\Pi_{g}$ & $-0.73$ & $-0.15$  & $-0.41$ & $-0.43$ & $ 0.23$  & $-0.28$ & $-0.51$ & $ 0.12$  & $-0.09$ & -$0.53$  & $-0.28$ \\
			\ce{S2    } & ${}^{3}\Sigma^{-}_{g}\rightarrow {}^{2}\Pi_{g}$ & $ 1.31$ & $ 1.62$  & $ 1.33$ & $ 1.42$ & $ 1.73$  & $ 1.34$ & $ 1.31$ & $ 1.64$  & $ 1.40$ &  $1.27$  & $1.33$  \\
			\ce{BO    } & ${}^{2}\Sigma^{+}\rightarrow {}^{1}\Sigma^{-}$  & $ 2.15$ & $ 2.65$  & $ 2.10$ & $ 2.09$ & $ 2.59$  & $ 2.30$ & $ 2.26$ & $ 2.76$  & $ 2.18$ &  $2.29$  & $2.36$  \\
			\ce{BN    } & ${}^{3}\Pi\rightarrow {}^{2}\Sigma^{+}$         & $ 2.92$ & $ 3.42$  & $ 2.46$ & $ 3.36$ & $ 3.79$  & $ 2.40$ & $ 2.92$ & $ 3.43$  & $ 2.49$ &  $2.71$  & $2.91$  \\
			\ce{SO    } & ${}^{3}\Sigma^{-}\rightarrow {}^{2}\Pi  $       & $ 0.62$ & $ 1.01$  & $ 0.84$ & $ 0.81$ & $ 1.22$  & $ 0.84$ & $ 0.63$ & $ 1.05$  & $ 1.00$ &  $0.69$  & $0.83$  \\ \hline
			\mae        &                                                 & $0.20$  &  $0.33$  & $0.18$  & $0.17$  &  $0.51$  & $0.14$  & $0.16$  &  $0.42$  & $0.16$  &  $0.12$  &         \\
			\std        &                                                 & $0.18$  &  $0.15$  & $0.15$  & $0.22$  &  $0.25$  & $0.16$  & $0.18$  &  $0.17$  & $0.18$  &  $0.06$  &         \\ \hline\hline
		\end{tabular}
	\end{threeparttable}
\end{table*}	

\begin{table*}[t!]
	\captionsetup{justification=raggedright,singlelinecheck=false,font=footnotesize}
	\caption{
		Vertical electron affinities (eV) of strongly spin contaminated open-shell molecules computed using EA-ADC, EA-EOM-CCSD, and CCSD(T) with various reference wavefunctions (UHF, OMP, and ROHF). 
		The aug-cc-pVDZ basis set was used in all calculations. 
		Also shown are mean absolute errors (\mae) and standard deviations (\std) relative to the CCSD(T) results. 
	}
	\label{tab:ADC_EA_strong}
	\setlength{\extrarowheight}{2pt}
	\setstretch{1}
	\tiny
	\centering
	\hspace*{-0.8cm}
	\begin{threeparttable}
		\begin{tabular}{llccccccccccc}
			\hline\hline
			System        & Transition                                       &  ADC(2)   & ADC(2)-X & ADC(3)  & ADC(2)  & ADC(2)-X & ADC(3)  & ADC(2)  & ADC(2)-X & ADC(3)  & EOM-CCSD & CCSD(T) \\
			              &                                                  &    UHF    &   UHF    &   UHF   & OMP(2)  &  OMP(2)  & OMP(3)  &  ROHF   &   ROHF   &  ROHF   &   UHF    &  ROHF   \\ \hline
			\ce{NO    }   & ${}^{2}\Pi \rightarrow {}^{3}\Sigma^{-}$         &  $-0.59$  & $-0.04$  & $-0.27$ & $-0.62$ & $-0.07$  & $-0.28$ & $-0.60$ & $-0.08$  & $-0.41$ & -$0.66$  & $-0.49$ \\
			\ce{NCO   }   & ${}^{2}\Pi\rightarrow {}^{1}\Sigma^{+}$          &  $ 3.41$  & $ 3.74$  & $ 2.95$ & $ 3.87$ & $ 3.78$  & $ 3.03$ & $ 3.82$ & $ 4.13$  & $ 3.05$ &  $3.14$  & $3.28$  \\
			\ce{AlO   }   & ${}^{2}\Sigma^{+}\rightarrow {}^{1}\Sigma^{+}$   &  $ 2.60$  & $ 3.46$  & $ 2.63$ & $ 3.26$ & $ 4.15$  & $ 2.60$ & $ 2.24$ & $ 2.65$  & $ 2.84$ &  $2.84$  & $2.51$  \\
			\ce{CNC   }   & ${}^{2}\Pi_{g} \rightarrow {}^{3}\Sigma^{-}_{g}$ &  $ 2.09$  & $ 2.42$  & $ 1.90$ & $ 2.26$ & $ 2.51$  & $ 1.86$ & $ 2.25$ & $ 2.47$  & $ 1.82$ &  $1.84$  & $1.87$  \\
			\ce{NO2   }   & ${}^{2}A_{1}\rightarrow {}^{1}A_{1}$             &  $ 1.35$  & $ 1.70$  & $ 1.22$ & $ 1.33$ & $ 1.76$  & $ 1.44$ & $ 1.41$ & $ 1.78$  & $ 1.27$ &  $1.21$  & $1.33$  \\
			\ce{CH2CHO}   & ${}^{2}A''\rightarrow {}^{1}A'$                  &  $ 1.26$  & $ 1.69$  & $ 1.16$ & $ 1.77$ & $ 2.14$  & $ 1.28$ & $ 1.63$ & $ 2.00$  & $ 1.31$ &  $1.24$  & $1.49$  \\
			\ce{C4O    }  & ${}^{3}\Sigma^{-}\rightarrow {}^{2}\Pi  $        & $   2.97$ & $ 3.20$  & $ 2.47$ & $ 3.19$ & $ 3.26$  & $ 2.56$ & $ 3.22$ & $ 3.37$  & $ 2.57$ & $ 2.60$  & $2.68$  \\
			\ce{BP}       & ${}^{3}\Pi\rightarrow {}^{4}\Sigma^{-}$          &  $ 2.55$  & $ 2.94$  & $ 2.37$ & $ 2.72$ & $ 3.05$  & $ 2.36$ & $ 2.63$ & $ 2.99$  & $ 2.39$ & $ 2.42$  & $2.53 $ \\
			\ce{C3H5    } & ${}^{2}A_{2}\rightarrow {}^{1}A_{1}$             & $  -0.03$ & $ 0.34$  & $-0.18$ & $ 0.52$ & $ 0.86$  & $ 0.05$ & $ 0.43$ & $ 0.76$  & $ 0.12$ & $-0.04$  & $0.20$  \\
			\ce{N3    }   & ${}^{2}\Pi_{g} \rightarrow {}^{1}\Sigma^{+}_{g}$ & $   2.85$ & $ 3.05$  & $ 2.01$ & $ 3.44$ & $ 3.45$  & $ 2.14$ & $ 3.14$ & $ 3.23$  & $ 2.31$ & $ 2.34$  & $2.52$  \\
			\ce{SCN   }   & ${}^{2}\Pi\rightarrow {}^{1}\Sigma^{+}$          &  $ 3.33$  & $ 3.66$  & $ 3.13$ & $ 3.71$ & $ 3.99$  & $ 3.27$ & $ 3.61$ & $ 3.93$  & $ 3.29$ &  $3.24$  & $3.35$  \\
			\ce{CH2CN }   & ${}^{2}B_{1}\rightarrow {}^{1}A_{1}$             &  $ 1.22$  & $ 1.58$  & $ 0.98$ & $ 1.60$ & $ 1.93$  & $ 1.17$ & $ 1.59$ & $ 1.93$  & $ 1.16$ &  $1.16$  & $1.33$  \\
			\ce{C2H3  }   & ${}^{2}A'\rightarrow {}^{1}A'$                   &  $-0.22$  & $ 0.28$  & $-0.28$ & $ 0.18$ & $ 0.64$  & $-0.14$ & $ 0.15$ & $ 0.61$  & $-0.15$ & -$0.14$  & $0.07$  \\
			\ce{CNN   }   & ${}^{3}\Sigma^{-}\rightarrow {}^{2}\Pi  $        &  $ 1.41$  & $ 1.86$  & $ 1.17$ & $ 2.15$ & $ 2.42$  & $ 1.31$ & $ 1.73$ & $ 2.14$  & $ 1.49$ &  $1.31$  & $1.51$  \\
			\ce{C3H3   }  & ${}^{2}B_{1}\rightarrow {}^{1}A_{1}$             & $   0.35$ & $ 0.70$  & $ 0.16$ & $ 0.80$ & $ 1.12$  & $ 0.37$ & $ 0.76$ & $ 1.10$  & $ 0.41$ & $ 0.32$  & $0.52$  \\
			\ce{CH    }   & ${}^{2}\Pi \rightarrow {}^{3}\Sigma^{-}$         &  $ 0.72$  & $ 1.43$  & $ 1.04$ & $ 1.01$ & $ 1.64$  & $ 1.14$ & $ 0.95$ & $ 1.57$  & $ 1.13$ &  $0.97$  & $1.08$  \\
			\ce{CHN2  }   & ${}^{2}A''\rightarrow {}^{1}A'$                  &  $ 1.62$  & $ 1.92$  & $ 1.08$ & $ 2.16$ & $ 2.24$  & $ 1.18$ & $ 1.97$ & $ 2.19$  & $ 1.34$ &  $1.30$  & $1.49$  \\
			\ce{HCCN  }   & ${}^{3}A''\rightarrow {}^{2}A''$                 &  $ 1.11$  & $ 1.59$  & $ 1.10$ & $ 1.71$ & $ 2.10$  & $ 1.32$ & $ 1.63$ & $ 2.07$  & $ 1.30$ &  $1.24$  & $1.46$  \\
			\ce{CN    }   & ${}^{2}\Sigma^{+}\rightarrow {}^{1}\Sigma^{+}$   &  $ 3.49$  & $ 4.08$  & $ 3.49$ & $ 3.60$ & $ 4.10$  & $ 3.76$ & $ 4.25$ & $ 4.62$  & $ 3.59$ &  $3.52$  & $3.69$  \\
			\ce{NS    }   & ${}^{2}\Pi\rightarrow {}^{3}\Sigma^{-}$          &  $ 1.32$  & $ 1.73$  & $ 1.44$ & $ 1.06$ & $ 1.34$  & $ 1.14$ & $ 1.24$ & $ 1.52$  & $ 1.03$ &  $0.94$  & $0.95$  \\
			\ce{SiC   }   & ${}^{3}\Pi\rightarrow {}^{2}\Sigma^{+}$          &  $ 2.05$  & $ 2.49$  & $ 1.97$ & $ 2.70$ & $ 2.96$  & $ 1.93$ & $ 2.34$ & $ 2.72$  & $ 1.94$ &  $2.03$  & $2.19$  \\
			\ce{CP    }   & ${}^{2}\Sigma^{+}\rightarrow {}^{1}\Sigma^{+}$   &  $ 2.23$  & $ 2.88$  & $ 2.08$ & $ 3.43$ & $ 3.54$  & $ 2.28$ & $ 3.69$ & $ 3.78$  & $ 2.17$ &  $2.39$  & $2.70$  \\ \hline
			\mae          &                                                  &  $0.20$   &  $0.38$  & $0.29$  & $0.37$  &  $0.66$  & $0.18$  & $0.31$  &  $0.60$  & $0.15$  &  $0.17$  &         \\
			\std          &                                                  &  $0.23$   &  $0.20$  & $0.25$  & $0.29$  &  $0.26$  & $0.17$  & $0.28$  &  $0.19$  & $0.16$  &  $0.13$  &         \\ \hline\hline
		\end{tabular}
	\end{threeparttable}
\end{table*}	

\begin{figure}[t!]
	\centering
	\captionsetup{justification=raggedright,singlelinecheck=false,font=footnotesize}		\includegraphics[scale=0.30,trim=0.0cm 0.0cm 0.0cm 0.0cm,clip]{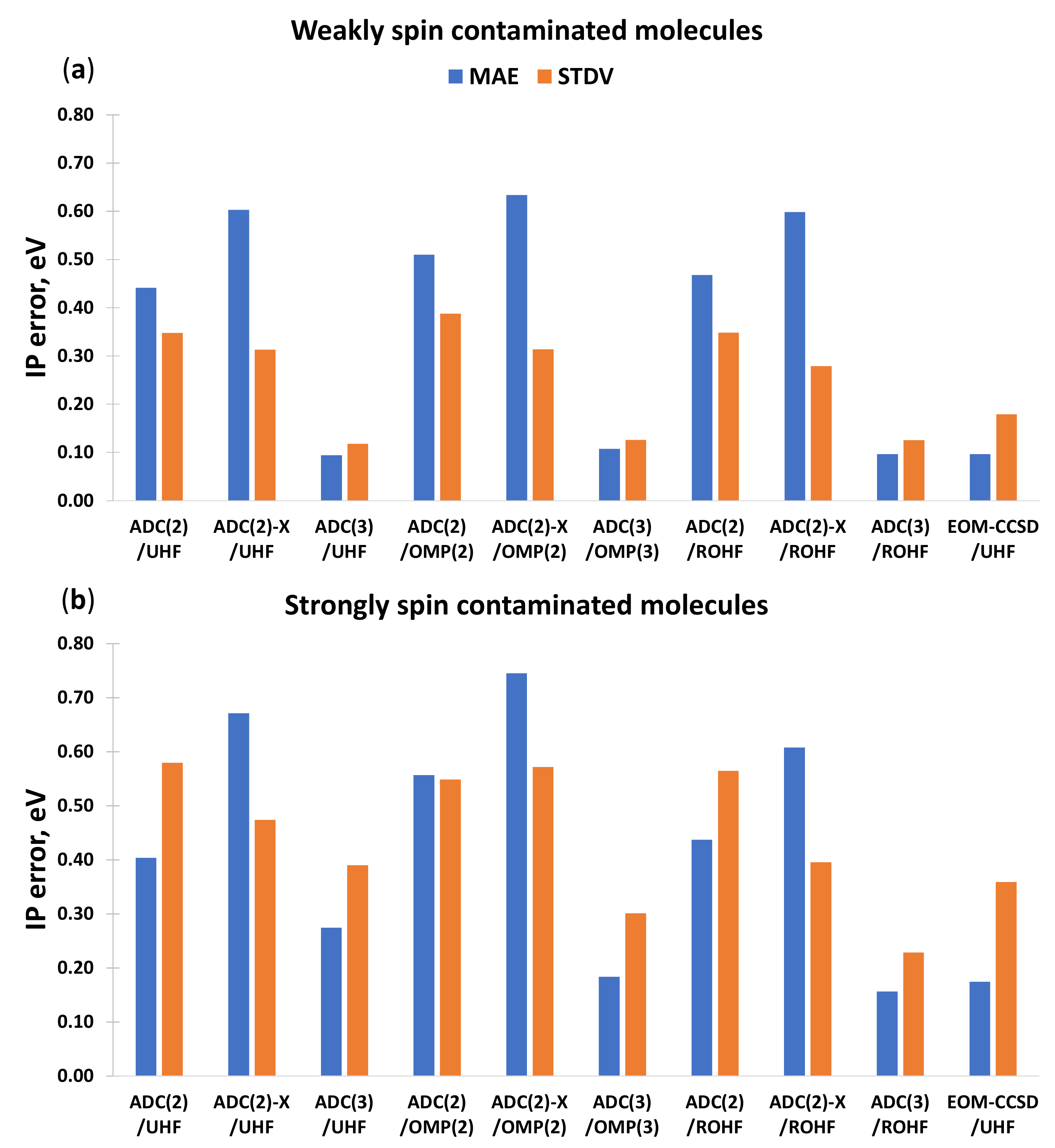}
	\caption{
		Mean absolute errors (MAE) and standard deviations (STDV) in the IP-ADC vertical ionization energies (eV) for (a) 18 weakly and (b) 22 strongly spin contaminated molecules with the ground-state UHF spin contamination of $< 0.1$ and $\ge 0.1$ a.u., respectively. 
		Reference data is from CCSD(T).
		The aug-cc-pVDZ basis set was used. 
		See \cref{tab:ADC_IP_weak,tab:ADC_IP_strong} for data on individual molecules.
	}
	\label{fig:ADC_IP}
\end{figure}

\begin{table*}[t!]
	\captionsetup{justification=raggedright,singlelinecheck=false,font=footnotesize}
	\caption{
		Vertical ionization energies (eV) of weakly spin contaminated open-shell molecules computed using IP-ADC, IP-EOM-CCSD, and CCSD(T) with various reference wavefunctions (UHF, OMP, and ROHF). 
		The aug-cc-pVDZ basis set was used in all calculations. 
		Also shown are mean absolute errors (\mae) and standard deviations (\std) relative to the CCSD(T) results. 
	}
	\label{tab:ADC_IP_weak}
	\setlength{\extrarowheight}{2pt}
	\setstretch{1}
	\tiny
	\centering
	\hspace*{-0.8cm}
	\begin{threeparttable}
		\begin{tabular}{llccccccccccc}
			\hline\hline
			System      & Transition                                       & ADC(2)  & ADC(2)-X & ADC(3)  & ADC(2)  & ADC(2)-X & ADC(3)  & ADC(2)  & ADC(2)-X & ADC(3)  & EOM-CCSD & CCSD(T) \\
			            &                                                  &   UHF   &   UHF    &   UHF   & OMP(2)  &  OMP(2)  & OMP(3)  &  ROHF   &   ROHF   &  ROHF   &   UHF    &  ROHF   \\ \hline
			\ce{BeH   } & ${}^{2}\Sigma^{+}\rightarrow {}^{1}\Sigma^{+}$   & $ 8.34$ & $ 8.18$  & $ 8.28$ & $ 8.31$ & $ 8.15$  & $ 8.25$ & $ 8.34$ & $ 8.18$  & $ 8.28$ & $ 8.32$  & $8.31 $ \\
			\ce{OH    } & ${}^{2}\Pi \rightarrow {}^{3}\Sigma^{-}$         & $11.79$ & $11.91$  & $13.03$ & $11.99$ & $12.16$  & $12.91$ & $11.76$ & $11.90$  & $13.05$ & $12.64$  & $12.75$ \\
			\ce{NH2   } & ${}^{2}B_{1}\rightarrow {}^{3}B_{1}$             & $11.15$ & $11.11$  & $11.95$ & $11.24$ & $11.23$  & $11.94$ & $11.14$ & $11.11$  & $11.97$ & $11.78$  & $11.85$ \\
			\ce{SH    } & ${}^{2}\Pi\rightarrow {}^{3}\Sigma^{-}$          & $ 9.79$ & $ 9.62$  & $ 9.99$ & $ 9.81$ & $ 9.65$  & $ 9.99$ & $ 9.79$ & $ 9.63$  & $ 9.99$ & $10.03$  & $10.01$ \\
			\ce{CH3   } & ${}^{2}A^{''}_{1}\rightarrow {}^{1}A^{'}_{1}$    & $ 9.40$ & $ 9.15$  & $ 9.64$ & $ 9.34$ & $ 9.10$  & $ 9.63$ & $ 9.34$ & $ 9.11$  & $ 9.61$ & $ 9.60$  & $9.64 $ \\
			\ce{SF    } & ${}^{2}\Pi\rightarrow {}^{3}\Sigma^{-}$          & $ 9.79$ & $ 9.70$  & $10.09$ & $ 9.75$ & $ 9.68$  & $10.13$ & $ 9.79$ & $ 9.70$  & $10.08$ & $10.19$  & $10.11$ \\
			\ce{OOH   } & ${}^{2}A''\rightarrow {}^{3}A''$                 & $10.70$ & $10.55$  & $11.74$ & $10.31$ & $10.23$  & $11.64$ & $10.70$ & $10.55$  & $11.76$ & $11.43$  & $11.40$ \\
			\ce{CH2   } & ${}^{3}B_{1}\rightarrow {}^{2}A_{1}$             & $10.11$ & $ 9.81$  & $10.27$ & $10.08$ & $ 9.80$  & $10.25$ & $10.07$ & $ 9.78$  & $10.26$ & $10.24$  & $10.27$ \\
			\ce{NH    } & ${}^{3}\Sigma^{-}\rightarrow {}^{2}\Pi  $        & $13.04$ & $12.68$  & $13.41$ & $13.07$ & $12.74$  & $13.38$ & $13.01$ & $12.66$  & $13.39$ & $13.29$  & $13.33$ \\
			\ce{PH2   } & ${}^{2}B_{1}\rightarrow {}^{1}A_{1}$             & $ 9.64$ & $ 9.29$  & $ 9.62$ & $ 9.58$ & $ 9.25$  & $ 9.61$ & $ 9.59$ & $ 9.26$  & $ 9.61$ & $ 9.67$  & $9.65 $ \\
			\ce{Si2   } & ${}^{3}\Sigma^{-}\rightarrow {}^{4}\Sigma^{-}  $ & $ 7.51$ & $ 7.20$  & $ 7.59$ & $ 7.35$ & $ 7.07$  & $ 7.55$ & $ 7.47$ & $ 7.18$  & $ 7.57$ & $ 7.74$  & $7.66 $ \\
			\ce{SiF   } & ${}^{2}\Pi \rightarrow {}^{1}\Sigma^{+}$         & $ 7.42$ & $ 7.20$  & $ 7.43$ & $ 7.32$ & $ 7.11$  & $ 7.42$ & $ 7.36$ & $ 7.16$  & $ 7.42$ & $ 7.52$  & $7.47 $ \\
			\ce{FO    } & ${}^{2}\Pi\rightarrow {}^{3}\Sigma^{-}$          & $11.88$ & $11.78$  & $12.96$ & $11.71$ & $11.73$  & $13.04$ & $11.89$ & $11.81$  & $12.97$ & $13.49$  & $12.82$ \\
			\ce{O2    } & ${}^{3}\Sigma^{-}_{g}\rightarrow {}^{2}\Pi_{g}$  & $11.26$ & $10.99$  & $12.39$ & $11.01$ & $10.88$  & $12.52$ & $11.30$ & $11.09$  & $12.41$ & $12.09$  & $12.24$ \\
			\ce{S2    } & ${}^{3}\Sigma^{-}_{g}\rightarrow {}^{2}\Pi_{g}$  & $ 8.95$ & $ 8.61$  & $ 9.33$ & $ 8.77$ & $ 8.52$  & $ 9.34$ & $ 8.91$ & $ 8.65$  & $ 9.31$ & $ 9.28$  & $9.27 $ \\
			\ce{BO    } & ${}^{2}\Sigma^{+}\rightarrow {}^{1}\Sigma$       & $12.61$ & $12.81$  & $13.17$ & $12.51$ & $12.50$  & $13.18$ & $12.63$ & $12.76$  & $13.04$ & $13.22$  & $12.95$ \\
			\ce{BN    } & ${}^{3}\Pi\rightarrow {}^{4}\Sigma^{-}$          & $10.49$ & $10.34$  & $11.25$ & $10.71$ & $10.71$  & $11.16$ & $10.57$ & $10.51$  & $11.23$ & $11.25$  & $11.28$ \\
			\ce{SO    } & ${}^{3}\Sigma^{-}\rightarrow {}^{2}\Pi  $        & $ 9.52$ & $ 9.45$  & $10.39$ & $ 9.22$ & $ 9.36$  & $10.44$ & $ 9.24$ & $ 9.46$  & $10.38$ & $10.28$  & $10.27$ \\ \hline
			\mae        &                                                  & $0.44$  &  $0.60$  & $0.09$  & $0.51$  &  $0.63$  & $0.11$  & $0.47$  &  $0.60$  & $0.10$  &  $0.10$  &         \\
			\std        &                                                  & $0.35$  &  $0.31$  & $0.12$  & $0.39$  &  $0.31$  & $0.13$  & $0.35$  &  $0.28$  & $0.13$  &  $0.18$  &         \\ \hline\hline
		\end{tabular}        
	\end{threeparttable}        
\end{table*}   	

\begin{table*}[t!]
	\captionsetup{justification=raggedright,singlelinecheck=false,font=footnotesize}
	\caption{
		Vertical ionization energies (eV) of strongly spin contaminated open-shell molecules computed using IP-ADC, IP-EOM-CCSD, and CCSD(T) with various reference wavefunctions (UHF, OMP, and ROHF). 
		The aug-cc-pVDZ basis set was used in all calculations. 
		Also shown are mean absolute errors (\mae) and standard deviations (\std) relative to the CCSD(T) results. 
	}
	\label{tab:ADC_IP_strong}
	\setlength{\extrarowheight}{2pt}
	\setstretch{1}
	\tiny
	\centering
	\hspace*{-0.8cm}
	\begin{threeparttable}
		\begin{tabular}{llccccccccccc}
			\hline\hline
			System        & Transition                                       &  ADC(2)   & ADC(2)-X &  ADC(3)  & ADC(2)  & ADC(2)-X & ADC(3)  & ADC(2)  & ADC(2)-X & ADC(3)  & EOM-CCSD & CCSD(T) \\
			              &                                                  &    UHF    &   UHF    &   UHF    & OMP(2)  &  OMP(2)  & OMP(3)  &  ROHF   &   ROHF   &  ROHF   &   UHF    &  ROHF   \\ \hline
			\ce{NO    }   & ${}^{2}\Pi \rightarrow {}^{1}\Sigma^{+}$         &  $ 8.92$  & $ 8.83$  & $ 9.71$  & $ 8.66$ & $ 8.68$  & $ 9.78$ & $ 8.94$ & $ 8.85$  & $ 9.56$ & $ 9.56$  & $9.54 $ \\
			\ce{NCO   }   & ${}^{2}\Pi\rightarrow {}^{3}\Sigma^{-}$          &  $11.14$  & $11.08$  & $11.57$  & $11.14$ & $11.20$  & $11.70$ & $11.22$ & $11.19$  & $11.61$ & $11.70$  & $11.69$ \\
			\ce{AlO   }   & ${}^{2}\Sigma^{+}\rightarrow {}^{1}\Sigma^{+}$   &  $ 9.00$  & $ 8.69$  & $ 9.05$  & $ 8.96$ & $ 8.74$  & $ 9.20$ & $ 7.98$ & $ 8.89$  & $ 9.89$ & $ 9.26$  & $9.88 $ \\
			\ce{CNC   }   & ${}^{2}\Pi_{g} \rightarrow {}^{1}\Sigma^{+}_{g}$ &  $ 9.72$  & $ 9.34$  & $ 9.66$  & $ 9.31$ & $ 9.03$  & $ 9.58$ & $ 9.55$ & $ 9.27$  & $ 9.52$ & $ 9.78$  & $9.54 $ \\
			\ce{NO2   }   & ${}^{2}A_{1}\rightarrow {}^{1}A_{1}$             &  $10.19$  & $10.19$  & $11.22$  & $ 9.70$ & $ 9.96$  & $11.53$ & $10.23$ & $10.33$  & $11.11$ & $11.23$  & $11.08$ \\
			\ce{CH2CHO}   & ${}^{2}A''\rightarrow {}^{1}A'$                  &  $ 9.97$  & $ 9.56$  & $10.25$  & $ 9.65$ & $ 9.41$  & $10.18$ & $ 9.77$ & $ 9.53$  & $10.11$ & $10.22$  & $10.16$ \\
			\ce{C4O    }  & ${}^{3}\Sigma^{-}\rightarrow {}^{2}\Pi  $        & $   9.52$ & $ 9.19$  & $ 9.56$  & $ 9.03$ & $ 8.96$  & $ 9.69$ & $ 9.42$ & $ 9.23$  & $ 9.41$ & $ 9.78$  & $9.63 $ \\
			\ce{BP}       & ${}^{3}\Pi\rightarrow {}^{4}\Sigma^{-}$          &  $ 8.91$  & $ 8.52$  & $ 8.94$  & $ 9.01$ & $ 8.79$  & $ 8.98$ & $ 8.91$ & $ 8.71$  & $ 8.96$ & $ 9.18$  & $9.16 $ \\
			\ce{C3H5    } & ${}^{2}A_{2}\rightarrow {}^{1}A_{1}$             & $   7.94$ & $ 7.51$  & $ 8.09$  & $ 7.43$ & $ 7.07$  & $ 7.91$ & $ 7.56$ & $ 7.22$  & $ 7.82$ & $ 8.05$  & $7.98 $ \\
			\ce{N3    }   & ${}^{2}\Pi_{g} \rightarrow {}^{3}\Sigma^{-}_{g}$ & $  10.32$ & $10.16$  & $ 10.36$ & $10.67$ & $10.69$  & $10.58$ & $10.62$ & $10.58$  & $10.38$ & $10.78$  & $10.75$ \\
			\ce{SCN   }   & ${}^{2}\Pi\rightarrow {}^{3}\Sigma^{-}$          &  $10.09$  & $ 9.88$  & $10.24$  & $10.09$ & $ 9.96$  & $10.37$ & $10.27$ & $10.08$  & $10.36$ & $10.45$  & $10.41$ \\
			\ce{CH2CN }   & ${}^{2}B_{1}\rightarrow {}^{1}A_{1}$             &  $10.04$  & $ 9.64$  & $10.20$  & $ 9.63$ & $ 9.32$  & $10.10$ & $ 9.92$ & $ 9.60$  & $ 9.99$ & $10.23$  & $10.09$ \\
			\ce{C2H3  }   & ${}^{2}A'\rightarrow {}^{1}A'$                   &  $ 9.64$  & $ 9.17$  & $ 9.84$  & $ 9.22$ & $ 8.77$  & $ 9.66$ & $ 9.39$ & $ 8.92$  & $ 9.64$ & $ 9.72$  & $9.61 $ \\
			\ce{CNN   }   & ${}^{3}\Sigma^{-}\rightarrow {}^{2}\Pi  $        &  $11.31$  & $10.93$  & $11.36$  & $10.80$ & $10.51$  & $11.25$ & $10.94$ & $10.54$  & $11.24$ & $11.40$  & $11.17$ \\
			\ce{C3H3   }  & ${}^{2}B_{1}\rightarrow {}^{1}A_{1}$             & $ 8.52  $ & $ 8.13$  & $ 8.67$  & $ 8.06$ & $ 7.72$  & $ 8.51$ & $ 8.31$ & $ 7.95$  & $ 8.43$ & $ 8.66$  & $8.53 $ \\
			\ce{CH    }   & ${}^{2}\Pi \rightarrow {}^{1}\Sigma^{+}$         &  $10.39$  & $ 9.95$  & $10.37$  & $10.43$ & $ 9.98$  & $10.40$ & $10.42$ & $ 9.96$  & $10.39$ & $10.46$  & $10.44$ \\
			\ce{CHN2  }   & ${}^{2}A''\rightarrow {}^{3}A''$                 &  $ 9.10$  & $ 8.76$  & $ 9.44$  & $ 9.54$ & $ 9.42$  & $ 9.62$ & $ 9.55$ & $ 9.29$  & $ 9.57$ & $ 9.76$  & $9.74 $ \\
			\ce{HCCN  }   & ${}^{3}A''\rightarrow {}^{2}A'$                  &  $10.49$  & $10.00$  & $10.65$  & $ 9.92$ & $ 9.64$  & $10.53$ & $10.31$ & $ 9.99$  & $10.36$ & $10.63$  & $10.48$ \\
			\ce{CN    }   & ${}^{2}\Sigma^{+}\rightarrow {}^{1}\Sigma$       &  $12.71$  & $12.74$  & $13.93$  & $12.81$ & $12.35$  & $14.27$ & $13.05$ & $13.13$  & $14.27$ & $13.84$  & $15.28$ \\
			\ce{NS    }   & ${}^{2}\Pi\rightarrow {}^{1}\Sigma^{+}$          &  $ 8.53$  & $ 8.18$  & $ 8.91$  & $ 7.72$ & $ 7.69$  & $ 8.94$ & $ 8.34$ & $ 8.06$  & $ 8.72$ & $ 8.79$  & $8.70 $ \\
			\ce{SiC   }   & ${}^{3}\Pi\rightarrow {}^{4}\Sigma^{-}$          &  $ 8.28$  & $ 7.98$  & $ 8.23$  & $ 8.56$ & $ 8.41$  & $ 8.54$ & $ 8.35$ & $ 8.11$  & $ 8.49$ & $ 8.68$  & $8.73 $ \\
			\ce{CP    }   & ${}^{2}\Sigma^{+}\rightarrow {}^{3}\Pi$          &  $10.45$  & $10.13$  & $10.39$  & $10.75$ & $10.62$  & $10.49$ & $10.66$ & $10.53$  & $10.42$ & $10.78$  & $10.74$ \\ \hline
			\mae          &                                                  &  $0.40$   &  $0.67$  &  $0.27$  & $0.56$  &  $0.74$  & $0.18$  & $0.44$  &  $0.61$  & $0.16$  &  $0.17$  &         \\
			\std          &                                                  &  $0.58$   &  $0.47$  &  $0.39$  & $0.55$  &  $0.57$  & $0.30$  & $0.56$  &  $0.40$  & $0.23$  &  $0.36$  &         \\ \hline\hline
		\end{tabular}        
	\end{threeparttable}        
\end{table*}   

We now turn our attention to the IP-ADC/UHF and IP-EOM-CCSD/UHF vertical ionization energies (IP's) of WSM and SSM presented in \cref{tab:ADC_IP_weak,tab:ADC_IP_strong}, respectively. 
The \mae and \std for each method computed relative to CCSD(T)/ROHF are depicted in \cref{fig:ADC_IP}.
As for EA-ADC, the largest \mae out of all IP-ADC/UHF approximations is demonstrated by IP-ADC(2)-X. 
For WSM, IP-ADC(3)/UHF is by far the most accurate UHF-based IP-ADC method, with \mae (0.09 eV) smaller than that of IP-ADC(2)/UHF (0.44 eV) by more than a factor of four. 
Upon increasing spin contamination from WSM to SSM, the IP-ADC/UHF calculations show trends similar to those in the EA-ADC/UHF results.
In particular, the \mae of IP-ADC(3)/UHF increases by a factor of three (from 0.09 to 0.27 eV), the IP-ADC(2)-X/UHF \mae grows by $\sim$ 12 \%, while IP-ADC(2)/UHF shows a small reduction in \mae from 0.44 to 0.40 eV.
All three IP-ADC/UHF methods also exhibit a very significant increase in \std, indicating that the growth in UHF spin contamination has a detrimental effect on reliability of these methods. 
The IP-EOM-CCSD/UHF method is similar to IP-ADC(3)/UHF in accuracy for WSM, but is significantly more accurate for SSM.

\begin{figure}[t!]
	\centering
	\captionsetup{justification=raggedright,singlelinecheck=false,font=footnotesize}
	\includegraphics[scale=0.30,trim=0.0cm 0.0cm 0.0cm 0.0cm,clip]{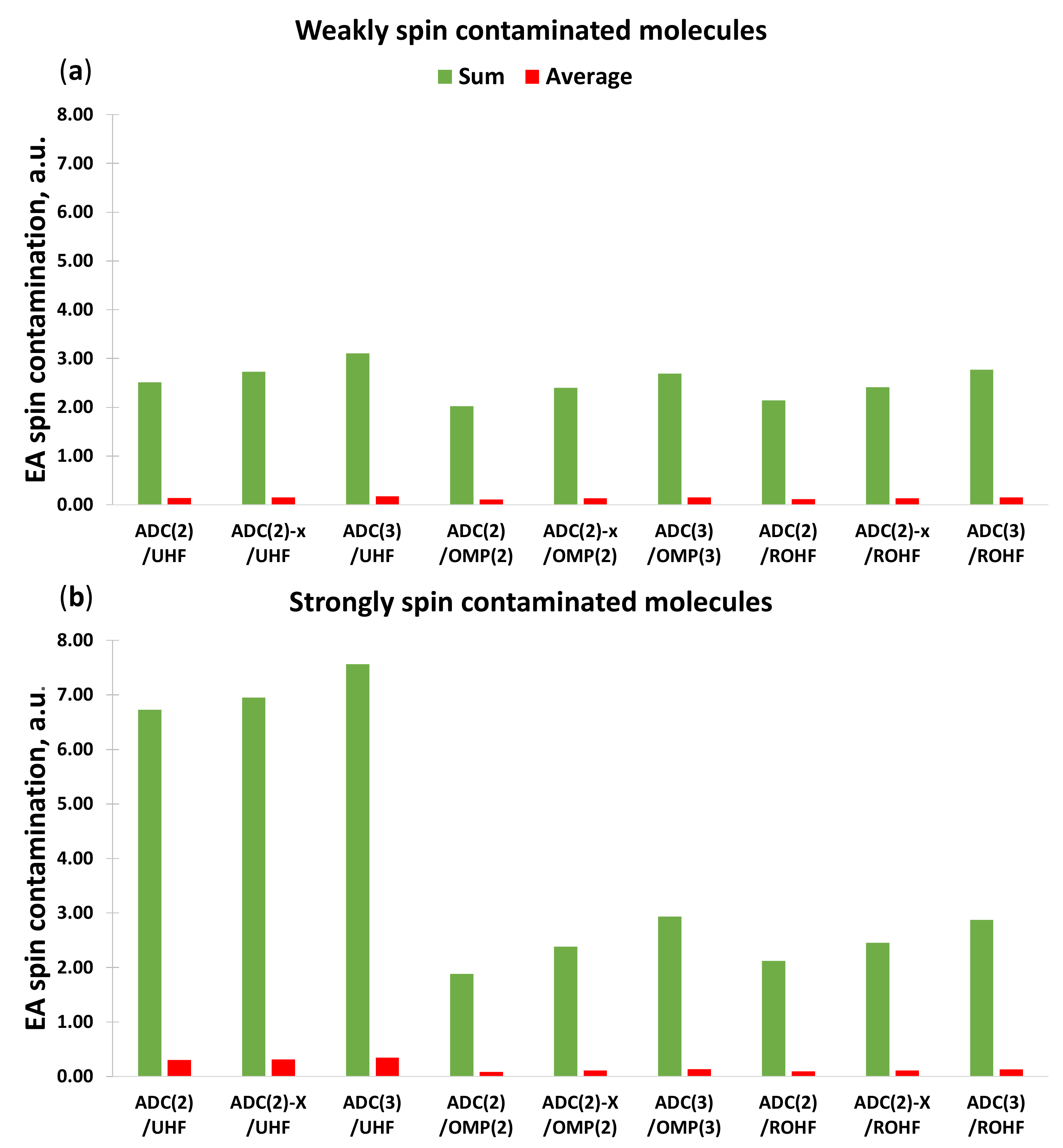}
	\caption{
		Sum and average of spin contamination in the lowest-energy electron-attached state for (a) 18 weakly and (b) 22 strongly spin contaminated molecules computed using EA-ADC with three different reference wavefunctions.
		The aug-cc-pVDZ basis set was used. 
	}
	\label{fig:ADC_EA_spin_c}
\end{figure}  

\begin{figure}[t!]
	\centering
	\captionsetup{justification=raggedright,singlelinecheck=false,font=footnotesize}
	\includegraphics[scale=0.30,trim=0.0cm 0.0cm 0.0cm 0.0cm,clip]{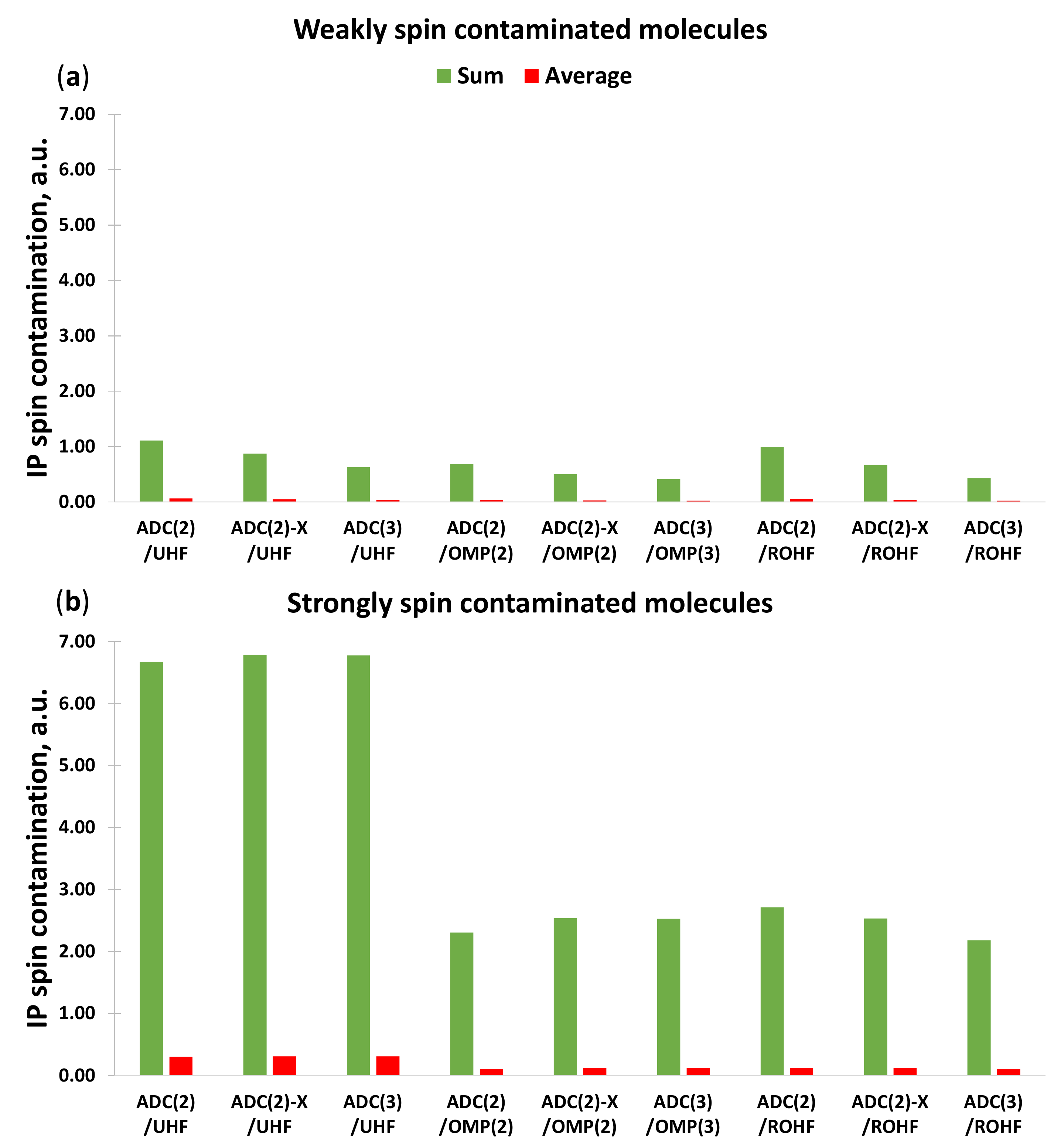}
	\caption{
		Sum and average of spin contamination in the lowest-energy ionized state for (a) 18 weakly and (b) 22 strongly spin contaminated molecules computed using IP-ADC with three different reference wavefunctions.
		The aug-cc-pVDZ basis set was used. 
	}
	\label{fig:ADC_IP_spin_c}
\end{figure}

To investigate how the performance of EA/IP-ADC/UHF methods is affected by spin contamination (\SC) in charged excited states, we computed the sum and average of \SC for WSM and SSM presented in \cref{fig:ADC_EA_spin_c,fig:ADC_IP_spin_c}.
The calculated \SC values for each molecule are tabulated in the Supplementary Information. 
For WSM, the \SC of IP-ADC/UHF decreases with the increasing level of IP-ADC theory and is smaller than the \SC in EA-ADC/UHF results.
The IP-ADC/UHF excited-state spin contamination is similar to the \SC in UHF reference (\cref{tab:ref_sc}) for all WSM, except for \ce{O2}, \ce{S2}, \ce{BO}, and \ce{SO} where the \SC of IP-ADC/UHF is significantly larger.
On the contrary, the excited-state \SC in EA-ADC/UHF grows with the increasing order of ADC approximation and is greater than the ground-state \SC from UHF for all WSM, with the exception of \ce{SiF}. 
Upon transitioning from WSM to SSM, the average \SC increases from $\sim$ 0.05 to $\sim$ 0.3 a.u.\@ for IP-ADC/UHF and from $\sim$ 0.15 to $\sim$ 0.34 a.u.\@ for EA-ADC/UHF.
In contrast to WSM, increasing the level of IP-ADC/UHF approximation for SSM does not lower the average \SC, which remains relatively constant ($\sim$ 0.3 a.u.). 
The EA-ADC/UHF results of SSM show a substantial increase in total and average \SC from EA-ADC(2) to EA-ADC(3). 

Overall, our results indicate that the increasing spin contamination in UHF reference significantly worsens the performance of EA/IP-ADC(2)-X/UHF and EA/IP-ADC(3)/UHF methods, while the accuracy of EA/IP-ADC(2)/UHF is affected much less.
However, when the UHF wavefunction is strongly spin-contaminated, charged excited states computed using all UHF-based ADC approximations exhibit a large spin contamination (\SC $>$ 0.1 a.u.).
In \cref{sec:results:OMPn,sec:results:ROHF}, we investigate how the performance of EA/IP-ADC approximations is affected by reducing the spin contamination in reference wavefunction.

\subsection{Reducing spin contamination using the orbital-optimized (OMP($n$)) reference}
\label{sec:results:OMPn}

Combining orbital optimization with conventional perturbation theories such as MP($n$) ($n$ = 2, 3) has been shown to significantly improve their accuracy for the open-shell molecules with significant spin contamination in the UHF reference wavefunction.\cite{Neese:2009p3060,bozkaya:2014p4389,soydas:2015p1564,bertels:2019p4170,soydas:2013p1452}
\cref{tab:ref_sc} demonstrates that incorporating the orbital relaxation and electron correlation effects together in OMP($n$) reduces the ground-state \SC to almost zero for all molecules studied in this work, while treating electron correlation with the UHF orbitals (MP($n$)/UHF) has a much smaller effect on spin contamination, resulting in a large \SC for most SSM.

\cref{fig:ADC_EA_spin_c,fig:ADC_IP_spin_c} depict the sum and average of \SC in the lowest-energy charged states of WSM and SSM computed using EA/IP-ADC($n$)/OMP($n$).
Optimizing the reference orbitals has a minor ($\sim$ 10\%) effect on the average \SC for WSM, but gives rise to a three-fold reduction in the average \SC for SSM.
The total and mean \SC of electron-attached states computed using EA-ADC($n$)/OMP($n$) have similar values for WSM and SSM at each order of perturbation theory, respectively (\cref{fig:ADC_EA_spin_c}).
For the ionized states, the average IP-ADC($n$)/OMP($n$) spin contamination for SSM remains significantly higher than that for WSM (\cref{fig:ADC_IP_spin_c}). 

\cref{fig:ADC_IP} illustrates that reducing the spin contamination in ground and excited states does not significantly affect the accuracy of IP-ADC for WSM (\cref{tab:ADC_IP_weak}).
For SSM, using the OMP($n$) orbitals increases the \mae and \std in IP-ADC(2) and IP-ADC(2)-X vertical ionization energies, while significantly lowering these error metrics for IP-ADC(3) (\cref{tab:ADC_IP_strong}).
The increase in IP-ADC(2)/OMP(2) \mae relative to IP-ADC(2)/UHF suggests that the smaller \mae of the UHF-based method is a result of fortuitous error cancellation and that reducing the spin contamination in IP-ADC(2)/OMP(2) shifts the balance of error worsening its performance.
For both WSM and SSM, IP-ADC(3)/OMP(3) shows nearly the same \mae (0.18 eV) and smaller \std (0.30 eV) relative to IP-EOM-CCSD/UHF, indicating that both methods are similarly accurate even for challenging open-shell molecules.

Similar trends are observed when comparing the errors in EA-ADC($n$) vertical electron affinities computed using the UHF and OMP($n$) reference orbitals, as illustrated in \cref{fig:ADC_EA}. 
Using the optimized orbitals does not significantly affect the accuracy of EA-ADC for WSM with the exception of EA-ADC(2)-X, which shows a large ($\sim$ 0.3 eV) increase in \mae (\cref{tab:ADC_EA_weak}).
When using an orbital-optimized reference for SSM, both EA-ADC(2) and EA-ADC(2)-X increase their \mae by $\sim$ 75 to 85 \% relative to the UHF-based methods (\cref{tab:ADC_EA_strong}).
Optimizing the orbitals for EA-ADC(3) shows a significant ($\sim$ 33 \%) reduction in the \mae for SSM that becomes nearly identical to that of EA-EOM-CCSD/UHF (0.17 eV).
Importantly, these changes in the relative performance of EA-ADC($n$) methods for SSM restore the expected trend \mae(EA-ADC(2)) $>$ \mae(EA-ADC(3)) that is not observed for the UHF reference.

To summarize, using the OMP($n$) reference helps to substantially lower the spin contamination in charged excited states of SSM computed using EA/IP-ADC($n$).
The reduction in \SC significantly improves the performance of EA/IP-ADC(3), which show \mae and \std similar to EA/IP-EOM-CCSD/UHF.
In contrast, using the optimized orbitals affects the balance of error cancellation in EA/IP-ADC(2) and EA/IP-ADC(2)-X increasing their errors.

\subsection{Reducing spin contamination using the restricted open-shell (ROHF) reference}
\label{sec:results:ROHF}

An alternative approach to reduce the spin contamination in post-Hartree--Fock calculations is to employ the ROHF reference (\cref{sec:theory:sr_adc_reference_wavefunctions}). 
We demonstrate this in \cref{tab:ref_sc}, which shows that the MP($n$)/ROHF ground-state spin contamination is close to zero for most of the open-shell molecules considered in this work.

\cref{fig:ADC_EA_spin_c,fig:ADC_IP_spin_c} present the excited-state \SC computed using EA/IP-ADC/ROHF.
As in \cref{sec:results:OMPn}, the largest differences with the EA/IP-ADC/UHF results are observed for SSM, where the ROHF-based EA/IP-ADC show much smaller excited-state spin contamination. 
The EA/IP-ADC with ROHF and OMP($n$) references exhibit similar mean \SC, although the former reference tends to produce somewhat larger spin contamination, as reflected by the sum of \SC across the charged excited states of all molecules. 

We now compare the performance of EA/IP-ADC methods with ROHF, OMP($n$), and UHF references.  
The \mae and \std of EA/IP-ADC($n$)/ROHF are quite similar to those computed using EA/IP-ADC($n$)/OMP($n$) as illustrated in \cref{fig:ADC_IP,fig:ADC_EA}.
Both ROHF and OMP(3) are equally effective in improving the accuracy of EA/IP-ADC(3) for SSM, with errors in vertical electron affinities and ionization energies similar to those from EA/IP-EOM-CCSD/UHF.
When combined with EA/IP-ADC(2) and EA/IP-ADC(2)-X, the \mae produced by the ROHF reference tend to be smaller than those from OMP(2) by $\sim$ 10 to 20 \%. 
This reduction in error is correlated with somewhat higher spin contamination observed in the EA/IP-ADC(2)/ROHF and EA/IP-ADC(2)-X/ROHF calculations in comparison to those computed using the OMP(2) reference and can be attributed to error cancellation as described in \cref{sec:results:OMPn}.

Overall, our results demonstrate that the ROHF reference is as effective as OMP($n$) in reducing the excited-state spin contamination and the errors of EA/IP-ADC(3) approximations. 
At the EA/IP-ADC(2) and EA/IP-ADC(2)-X levels of theory, the ROHF calculations exhibit smaller errors compared to OMP($n$), but may suffer from somewhat higher spin contamination. 

\section{Conclusions}
\label{sec:conclusions}

In this work, we investigated the effect of spin contamination on the performance of three single-reference ADC methods for the charged excitations of open-shell systems (EA/IP-ADC(2), EA/IP-ADC(2)-X, and EA/IP-ADC(3)).
To this end, we benchmarked the accuracy of EA/IP-ADC for 40 molecules with different levels of spin contamination in the unrestricted Hartree--Fock (UHF) reference wavefunction and developed an approach for calculating the expectation values of spin-squared operator and spin contamination in the EA/IP-ADC charged excited states. 
Our study demonstrates that the EA/IP-ADC results can be affected by significant spin contamination (\SC), especially when \SC in the UHF reference wavefunction is large ($\ge$ 0.1 a.u.).
For such strongly spin contaminated systems, the average errors of third-order ADC approximations (EA/IP-ADC(3)/UHF) increase by $\sim$ 60 \% for EA and $\sim$ 300 \% for IP, relative to molecules with the UHF \SC $<$ 0.1 a.u. 
The extended second-order methods (EA/IP-ADC(2)-X/UHF) also show significant increase (10 to 15 \%) in their average errors upon increasing the spin contamination in UHF reference, while the accuracy of strict second-order approximations (EA/IP-ADC(2)/UHF) is affected much less.

To mitigate the spin contamination in ADC calculations, we implemented the EA/IP-ADC methods with reference orbitals from restricted open-shell Hartree--Fock (ROHF) and orbital-optimized $n$th-order M\o ller--Plesset perturbation (OMP($n$)) theories.
The results of our work provide a clear evidence that the accuracy of EA/IP-ADC(3) is quite sensitive to the spin contamination in reference wavefunction.
For strongly spin contaminated open-shell molecules (\SC $\ge$ 0.1 a.u.), combining EA/IP-ADC(3) with ROHF or OMP(3) increases their accuracy by $\sim$ 30 to 50 \%.
While both ROHF and OMP(3) are equally effective in reducing spin contamination and improving the performance of EA/IP-ADC(3), the ROHF reference has a much lower computational scaling with the basis set size ($\mathcal{O}(N^4)$) relative to that of OMP(3) ($\mathcal{O}(N^6)$).

For EA/IP-ADC(2) and EA/IP-ADC(2)-X, using the ROHF or OMP(2) orbitals reduces the spin contamination in ground and excited electronic states of open-shell molecules, but increases the errors in vertical electron affinities and ionization energies.
Although employing ROHF or OMP(2) leads to a significant loss in EA/IP-ADC(2) and EA/IP-ADC(2)-X accuracy for calculating the charged excitation energies, these reference wavefunctions may still be preferred over UHF if one is interested in calculating other excited-state properties that can be sensitive to spin contamination.

\section{Supplementary Material}
See the supplementary material for the working equations of EA/IP-ADC($n$) with the ROHF reference, equations for the one- and two-particle density matrices, Cartesian geometries of 40 neutral open-shell molecules, as well as excited-state spin contamination and spectroscopic factors computed using all ADC methods. 

\section{Acknowledgements}
This work was supported by the start-up funds from the Ohio State University and by the National Science Foundation, under Grant No.\@ CHE-2044648. 
Additionally, S.B. was supported by a fellowship from Molecular Sciences Software Institute under NSF Grant No.\@ ACI-1547580.
Computations were performed at the Ohio Supercomputer Center under projects PAS1583 and PAS1963.\cite{OhioSupercomputerCenter1987} 
The authors thank Ruojing Peng for contributions in the initial stage of this project.

\section{Data availability}
The data that supports the findings of this study are available within the article and its supplementary material. Additional data can be made available upon reasonable request.


\end{document}